\documentclass[aps,preprint]{revtex4}
\usepackage[T1]{fontenc}
\usepackage[utf8]{inputenc}
\usepackage{array}
\usepackage{multirow}
\usepackage{amsmath}
\usepackage{amssymb}
\usepackage{graphicx}
\usepackage{makecell}
\usepackage{color} 

\makeatletter

\providecommand{\tabularnewline}{\\}

\@ifundefined{textcolor}{}
{%
 \definecolor{BLACK}{gray}{0}
 \definecolor{WHITE}{gray}{1}
 \definecolor{RED}{rgb}{1,0,0}
 \definecolor{GREEN}{rgb}{0,1,0}
 \definecolor{BLUE}{rgb}{0,0,1}
 \definecolor{CYAN}{cmyk}{1,0,0,0}
 \definecolor{MAGENTA}{cmyk}{0,1,0,0}
 \definecolor{YELLOW}{cmyk}{0,0,1,0}
}

\usepackage{amsfonts}
\usepackage{array}
\usepackage{multirow}
\usepackage{latexsym}
\usepackage{epsfig}
\usepackage{float}
\usepackage{dsfont}

\usepackage{ragged2e}
\justifying\let\raggedright\justifying
\usepackage{caption}
\providecommand{\tabularnewline}{\\}

\newcommand{\be}{\begin{equation}}
\newcommand{\ee}{\end{equation}}

\RequirePackage{doi}

\makeatother

\begin{document}
\title{Resonant Graviton-Photon Conversion with Stochastic Magnetic Field
in the Expanding Universe}
\author{Andrea Addazi}
\affiliation{Center for Theoretical Physics, College of Physics, Sichuan University,
Chengdu, 610064, PR China}
\affiliation{INFN, Laboratori Nazionali di Frascati, Via E. Fermi 54, I-00044 Roma,
Italy}
\author{Salvatore Capozziello}
\affiliation{Dipartimento di Fisica ”E. Pancini”, Universita di Napoli “Federico
II”, and Istituto Nazionale di Fisica Nucleare, Sezione di Napoli,
Compl. Univ. di Monte S. Angelo, Edificio G, Via Cinthia, I-80126,
Napoli, Italy}
\affiliation{Scuola Superiore Meridionale, Largo S. Marcellino 10, I-80138, Napoli,
Italy}
\author{Qingyu Gan}
\affiliation{Scuola Superiore Meridionale, Largo S. Marcellino 10, I-80138, Napoli,
Italy}
\begin{abstract}
We investigate graviton-photon oscillations sourced by cosmological magnetic fields from Gertsenshtein effect. 
We adopt
a robust perturbative approach and we find that the conversion probability
from graviton to photon can be resonantly enhanced in monochromatic, multi-chromatic
and scale invariant spectrum models of stochastic magnetic field fluctuations.
In addition, the expansion of the Universe acts as a decoherence factor,
which demands a natural discretization scheme along the line of sight.
Including also decoherence from cosmic acceleration, 
we find that conversion probabilities for stochastic magnetic fields 
are completely different than results predicted from existing magnetic domain-like
models in a wide range of magnetic strengths and coherence lengths.
Resonances can be tested by radio telescopes as a probe of high frequency
gravitational wave sources and primordial magnetogenesis mechanisms.
\end{abstract}
\maketitle
\tableofcontents{}

\section{Introduction}

In last decade, a large amount of astrophysical and cosmological data in electromagnetic (EM) and gravitational wave (GW)
channels opens the possibility to test several new physics mechanisms in the early Universe from the multi-messenger approach. 
 Intriguingly, EM and GW
radiations can be entwined through the (inverse) Gertsenshtein effect.
This describes a conversion between graviton and photon in the presence
of a magnetic background field \citep{gertsenshtein1962wave}. Such an 
effect can have several cosmological implications such as 
detecting the GW signals via radio-wave channel and vice
versa \citep{Cillis:1996qy,Caprini:2006jb,Pshirkov:2009sf,Capozziello:2018qjs,Khodadi:2018scn,Domcke:2020yzq,Fujita:2020rdx},
probing the primordial magnetic field \citep{Chen:2013gva}, testing QED corrections \citep{chen1995resonant,Dolgov:2012be,Dolgov:2013pwa},
constraining the dark photons \citep{Masaki:2018eut}, testing
the modified gravity theory \citep{Capozziello:2022dle,Cembranos:2023ere}
and so on. In particular, in Ref. \citep{Domcke:2020yzq} the authors
obtained the existing strongest upper bounds on the MHz-GHz primordial
GWs derived from radio telescopes ARCADE 2 \citep{fixsen2011arcade}
and EDGES \citep{Bowman:2018yin}.

The magnetic background field plays a pivotal role in the Gertsenshtein
mechanism. Concerning the cosmic magnetic field in intergalactic scale,
its origin remains a mystery yet.
The observational bounds on the strength and coherence
length of intergalactic magnetic fields are obtained from Faraday
rotation measures \citep{kronberg1994extragalactic}, CMB analysis
\citep{Planck:2015zrl}, magnetohydrodynamics processes \citep{Jedamzik:1996wp,Kahniashvili:2012uj}
and blazar gamma-rays \citep{taylor2011extragalactic}. 
 Theoretical considerations point
to two typical magnetogenesis scenarios in the primordial Universe:
cosmological first order phase transitions and inflation. For phase transitions, causality bounds lead to a blue spectrum of the
magnetic field. On the other hand, inflation models can generate a scale invariant spectrum. 
(See Ref. \citep{Durrer:2013pga,Subramanian:2015lua} for a comprehensive
review on these subjects). 

The graviton-photon mixing is described by 
a system of equations which can be 
 analytically solved for a constant
magnetic field \citep{Ejlli:2020fpt}. 
For the more complicated case of 
an inhomogeneous
magnetic background, two main approaches have been proposed. The first, known as the ``domain-like' model
\citep{Cillis:1996qy,Deffayet:2001pc,Grossman:2002by,Mirizzi:2005ng,Pshirkov:2009sf,Bassan:2010ya,Meyer:2014epa,Evoli:2016zhj,Masaki:2017aea,Galanti:2018nvl,Schiavone:2021imu,Kachelriess:2021rzc},
assumes that the line-of-sight path can be divided into many patches with size 
of about the coherence length of the magnetic field. In each domain, the magnetic
field is assumed to be uniform but its direction is randomly chosen.
This approach is widely adopted for simplicity.
On the contrary, the stochastic approach assumes that
magnetic random perturbations are isotropic and Gaussian distributed. 
In the statistical approach,   one can perform perturbative methods,
starting with the power spectrum of the stochastic magnetic field and
solving wave equations for perturbations around it \citep{raffelt1988mixing,Mirizzi:2007hr,Meyer:2014epa,Evoli:2016zhj,Domcke:2020yzq,carenza2021turbulent,Marsh:2021ajy}.
Within this framework, in Refs. \citep{Domcke:2020yzq,carenza2021turbulent}
the authors found an interesting phenomenon: graviton-photon conversion probability
has a linear growth with the propagation distance. In fact,
this is a parametric resonance arising from a magnetic oscillation
in certain frequency which was firstly addressed in a photon-axion mixing
system \citep{raffelt1988mixing}. 
Moreover, similar linear amplification of the conversion probability
has also been found from inhomogeneity of electron density
perturbations \citep{carlson1994photon}. Despite its significance,
graviton-photon resonances have not received a wide attention in litterature. 


As mentioned above, in most of domain-like models the mean
patch size is assumed to be of the order of the coherence length of the magnetic field.
This can be theoretically conceivable in
the causal magnetogenesis from phase transition.
However,
for magnetogenesis from inflation,
in the case of a scale-invariant spectrum, the coherence length loses
its clear physical interpretation rendering the above discretization
scheme inadequate. Moreover, non-commutativity
of the wave equation at different cosmic time in an expanding Universe
leads to a type of decoherence effect which is commonly neglected in domain-like approaches.
In general the decoherence factors for graviton-photon oscillations (e.g.,
the inhomogeneity of the electron density, inelastic scattering of
the participating particles and so on) demands a proper discretization
scheme along the line of sight.


In this paper, we will explore parametric resonances 
in graviton-photon transitions with cosmological stochastic magnetic fields. 
As mentioned above such a phenomenon was poorly analysed in litterature. 
We will approach the problem with perturbative methods 
including decoherence from cosmic acceleration
with discretization scheme beyond domain-like one. 
In our analysis, we consider both mono-chromatic and 
scale-invariant power spectra of random magnetic field fluctuations. 
Remarkably, we will show that, in several regions of parameters,
resonances can enhance graviton-photon oscillation probability 
up to $2\sim5$ orders of magnitude with respect to results given in Ref. \citep{Domcke:2020yzq}.
Furthermore,  resonance effects, with their distinct
features,  in relation to the magnetic spectrum,
can be employed as valuable probes to discern the magnetogenesis models in
 early Universe.

Our paper is organized as follows. In Sec. \ref{sec:Gertsenshtein-Effect},
we briefly introduce the graviton-photon mixing system in the static
Universe and obtain the expression for the conversion probability
in inhomogeneous magnetic background using the perturbative approach.
In Sec. \ref{sec:Resonance}, we show the parametric resonances
of two representative power spectra of magnetic field. In order to account
for the expansion of the Universe, we apply the discretization scheme
based on steady approximation and show the numerical results in
Sec. \ref{sec:Discritization-scheme}. In Sec. \ref{sec:Discussion-and-Conclusion}
we show our conclusions, discussions and remarks on our results and future prospectives. 

\section{The Gertsenshtein Effect}

\label{sec:Gertsenshtein-Effect}
The (inverse) Gertsenshtein effect describes the conversion between
the graviton and photon in the presence of a magnetic field.  As a first approximation, we assume
Universe expansion as negligible and consider graviton-photon transitions in Minkowski spacetime. Let us consider gravitons propagate along the $l$-direction and convert to photons (we avoid using “$z$-direction” as $z$ denotes the redshift of Universe). 
As shown in Ref.  \citep{Ejlli:2018hke}, the mixing system including two polarization states of photon as well as graviton in an external magnetic background $\mathbf{B}=(\mathbf{B}_x(l),\mathbf{B}_y(l),\mathbf{B}_z(l))$  is described by 
\begin{equation}
\left(\omega+i \partial_l\right) \boldsymbol{I}_{4 \times 4}\left(\begin{array}{c}
		h_{\times}(\omega,l) \\
		h_{+}(\omega,l) \\
		A_x(\omega,l) \\
		A_y(\omega,l)
	\end{array}\right)+\left(\begin{array}{cccc}
		0 & 0 & -i M_{g \gamma}^x & i M_{g \gamma}^y \\
		0 & 0 & i M_{g \gamma}^y & i M_{g \gamma}^x \\
		i M_{g \gamma}^x & -i M_{g \gamma}^y & M_x & M_{\mathrm{CF}} \\
		-i M_{g \gamma}^y & -i M_{g \gamma}^x & M_{\mathrm{CF}}^* & M_y
	\end{array}\right)\left(\begin{array}{c}
		h_{\times}(\omega,l) \\
		h_{+}(\omega,l) \\
		A_x(\omega,l) \\
		A_y(\omega,l)
	\end{array}\right)=0 \label{eq:eom1}
\end{equation}
with the mixing matrix elements   
\begin{equation}
	M_{g \gamma}^x  =\frac{1}{2}\kappa  \mathbf{B}_x, \quad
	M_{g \gamma}^y  =\frac{1}{2} \kappa  \mathbf{B}_y, \quad
	M_x  =-\frac{1}{2}\frac{\Pi_{x x} }{ \omega}, \quad
	M_y  =-\frac{1}{2}\frac{\Pi_{y y} }{ \omega}, \quad
	M_{\mathrm{CF}}  =-\frac{1}{2}\frac{\Pi_{x y} }{ \omega}. \label{M-Pi}
\end{equation}
We work in Planck unit and set  $\kappa=\sqrt{16 \pi}$. Here, $A_{x/y}(\omega,l)$ and  $h_{+/\times}(\omega,l)$ are polarized modes of photons and gravitons in frequency $\omega$, respectively.  We point out that above formula is only valid when  the external magnetic field 
varies in space on much larger
scales than  the wavelength of mixing gravitons as well as photons. Although the cosmic magnetic field has not been observed yet,  theoretical constraints suggest that it coherently oscillates at a certain scale larger than $1\textrm{pc}$ \citep{Durrer:2013pga}. It indicates that Eq. (\ref{eq:eom1}) can be  a good approximation since the frequency of the mixing system of our interests is typically $\omega\gtrsim\textrm{MHz}$ (corresponding to the wavelength $\lesssim \textrm{km}$).

Concerning the $\Pi_{xx/xy/yy}$ terms, they present the effects of the uniform plasma as well as transverse magnetic background on the polarized photons. As shown in \cite{Ejlli:2016asd,Ejlli:2018hke}, the Cotton-Mouton effect due to the external magnetic field can be schematically captured in $\sim \frac{\omega_{\mathrm{pl}}^2  \omega_c^2}{\omega^2-\omega_c^2}$ with plasma frequency $\omega_{\mathrm{pl}}=\sqrt{ e^2 n_e / m_e}$ and cyclotron frequency $\omega_c=e B / m_e$. Here $m_e$, $e$ and $n_e$ are respectively the mass, charge and number density of the free electrons in the plasma and $B=|\mathbf{B}|$ is the strength of the magnetic field.  In our setup with $\omega\gtrsim\textrm{MHz}$ and $B \lesssim 10^{-9} \textrm{Gauss}$, the typical scales of plasma and cyclotron frequencies are $\omega_{\mathrm{pl}} \simeq 1 \textrm{Hz} $ and $\omega_c \lesssim 0.01 \textrm{Hz}$, which indicates that the Cotton-Mouton effect has little contribution to $\Pi_{xx/xy/yy}$ terms. In addition, the  Faraday Rotation included in  $\Pi_{xy}$ term  describes the coupling between two photon polarization states. It plays the role in the analysis of the conversion between two polarized modes of photons, which is irrelevant for the
problem at hand. In short, we can safely neglect the Cotton-Mouton effect as well as the Faraday Rotation and only  include the plasma effect as  $\Pi_{x x}\simeq \Pi_{yy}  \simeq \omega_{\mathrm{pl}}^2 $ and  $\Pi_{x y}\simeq  0 $. 
Then from Eq. (\ref{M-Pi}),  Eq. (\ref{eq:eom1}) can be simplified as 
\begin{equation}
	\begin{aligned}	
	&	\partial_l\left(\begin{array}{c}
			h_{\times}(\omega, l) \\
			h_{+}(\omega, l) \\
			A_x(\omega, l) \\
			A_y(\omega, l)
		\end{array}\right)  =i K(l)\left(\begin{array}{c}
			h_{\times}(\omega, l) \\
			h_{+}(\omega, l) \\
			A_x(\omega, l) \\
			A_y(\omega, l)
		\end{array}\right), \\
 &K(l)=\left(\begin{array}{cccc}
			\omega & 0 & -i \frac{1}{2} \kappa \mathbf{B}_x (l) & i \frac{1}{2} \kappa \mathbf{B}_y(l) \\
			0 & \omega & i \frac{1}{2} \kappa \mathbf{B}_y(l) & i \frac{1}{2} \kappa \mathbf{B}_x(l) \\
			i \frac{1}{2} \kappa \mathbf{B}_x(l) & -i \frac{1}{2} \kappa \mathbf{B}_y(l) & \omega\left(1+n_{\mathrm{pl}}\right) & 0 \\
			-i \frac{1}{2} \kappa \mathbf{B}_y(l) & -i \frac{1}{2} \kappa \mathbf{B}_x(l) & 0 & \omega\left(1+n_{\mathrm{pl}}\right)
		\end{array}\right), \label{eom2}
	\end{aligned} 
\end{equation}
where we introduce the refraction index $n_\textrm{pl} = -\omega_\textrm{pl}^2/(2 \omega^2)$. To obtain the conversion probability in the graviton-photon mixing, we
introduce a conversion matrix $\mathcal{U}$ in 4$\times$4 size given by $(h_{\times}(l),h_{+}(l),A_x(l),A_y(l))^T= \mathcal{U}(l,l_0) (h_{\times}(l_0),h_{+}(l_0),A_x(l_0),A_y(l_0))^T$. The conversion matrix $\mathcal{U}\left(l, l_0\right)$ has a clear physical interpretation as each component represents the corresponding conversion process among different polarized modes of gravitons and photons.
Let us concentrate on the transition process from the gravitons to photons and assume that the GW source to be unpolarized with source's intensity normalized to unity. Thus the initial conditions are chosen to be $A_x\left(l_0\right)=A_y\left(l_0\right)=0$,  $  h_{\times}\left(l_0\right)=c_1$ and $h_{+}\left(l_0\right)=c_2$, where the random complex constants statistically average to zero but their mean squares remain nonvanishing, i.e.,  $\left\langle c_1 \right\rangle =\left\langle c_2 \right\rangle =0$ and  $\left\langle |c_1|^2 \right\rangle =\left\langle |c_2|^2 \right\rangle =1/2$. At distance $l$, the generating photons can
be easily read from the conversion matrix, namely $ A_x(l)=\mathcal{U}_{31} c_1+\mathcal{U}_{32} c_2$ and $  A_y(l)=\mathcal{U}_{41} c_1+\mathcal{U}_{42} c_2$. Performing an average over the initial unpolarized GW state, we obtain the conversion probability 
\begin{equation}
	\left\langle \mathcal{P}(l) \right\rangle	=  \frac{\left\langle\left|A_x(l)\right|^2 \right\rangle+\left\langle \left|A_y(l)\right|^2 \right\rangle}{\left\langle\left|h_{\times}\left(l_0\right)\right|^2\right\rangle+\left\langle\left|h_{+}\left(l_0\right)\right|^2\right\rangle} 
		=  \frac{1}{2}\left(\left|\mathcal{U}_{31}\right|^2+\left|\mathcal{U}_{32}\right|^2+\left|\mathcal{U}_{41}\right|^2+\left|\mathcal{U}_{42}\right|^2 \right), \label{P}
\end{equation} 
where the interference terms (e.g., $c_1c^*_2\mathcal{U}_{31} \mathcal{U}_{32}^*$  etc.) vanish due to $\left\langle c_1 \right\rangle =\left\langle c_2 \right\rangle =0$.

To compute conversion matrix  $\mathcal{U}$, we rewrite Eq. (\ref{eom2}) in form of
\begin{equation}
	\partial_l \mathcal{U}\left(l, l_0\right)=i K(l) \mathcal{U}\left(l, l_0\right). \label{eom-U}
\end{equation} We firstly consider the mixing matrix $K$ to be $l$-independence, which allows us to express
the exact solution in a closed form  $\mathcal{U}(l,l_{0})=\textrm{exp}\left[i\left(l-l_{0}\right)K\right]$. One can directly calculate $\left|\mathcal{U}_{31}\right|^2=\left|\mathcal{U}_{42}\right|^2=\frac{1}{4} \kappa^2 \mathbf{B}_x^2 l_{\mathrm{osc}}^2 \sin ^2\left((l-l_0) / l_{\mathrm{osc}}\right)  $ and $\left|\mathcal{U}_{32}\right|^2=\left|\mathcal{U}_{41}\right|^2=\frac{1}{4} \kappa^2 \mathbf{B}_y^2 l_{\mathrm{osc}}^2 \sin ^2\left((l-l_0) / l_{\mathrm{osc}}\right)$, 
then obtain the probability of the graviton transition to photon for a traveling
 distance $\triangle l=l-l_0$ as
\begin{equation}
\left\langle \mathcal{P}(\triangle l) \right\rangle   =\frac{1}{4}\kappa^{2}(\mathbf{B}_x^2+\mathbf{B}_y^2)l_{\textrm{osc}}^{2}\sin^{2}\left(\triangle l/l_{\textrm{osc}}\right).\label{eq:P-constantB}
\end{equation}
Here $l_{\textrm{osc}}=2/\sqrt{\kappa^{2}(\mathbf{B}_x^2+\mathbf{B}_y^2)+n_{\textrm{pl}}^{2}\omega^{2}}$
is the typical length scale of graviton-photon mixing 
oscillations.

Regarding the general case with a spatial varying magnetic field distributed along the path of the GW propagation, the non-commutativity of $\left[K(l),K(l')\right]\neq0$
prevents us from expressing $\mathcal{U}(l,l_{0})$ in an exponential
form as the case of constant magnetic field. Since  the cosmic magnetic field is expected to be significantly suppressed due to the high isotropy and homogeneity of our Universe, the coupling between the graviton-photon mixing system and the external magnetic field can be treated as a perturbative interaction.
 We follow the same perturbative approach in Refs.
\citep{raffelt1988mixing,Mirizzi:2007hr,Evoli:2016zhj,Ejlli:2018hke,Domcke:2020yzq} and split
$K(l)$ as
$K(l)  =  K_{0}+\delta K(l)$ with
\begin{equation}
K_0=\left(\begin{array}{cccc}
	\omega & 0 & 0 & 0 \\
	0 & \omega & 0 & 0 \\
	0 & 0 & \omega\left(n_{\mathrm{pl}}+1\right) & 0 \\
	0 & 0 & 0 & \omega\left(n_{\mathrm{pl}}+1\right)
\end{array}\right),
\delta K(l)=\left(\begin{array}{cccc}
	0 & 0 & -i \frac{1}{2} \kappa \mathbf{B}_x & i \frac{1}{2} \kappa \mathbf{B}_y \\
	0 & 0 & i \frac{1}{2} \kappa \mathbf{B}_y & i \frac{1}{2} \kappa \mathbf{B}_x \\
	i \frac{1}{2} \kappa \mathbf{B}_x & -i \frac{1}{2} \kappa \mathbf{B}_y & 0 & 0 \\
	-i \frac{1}{2} \kappa \mathbf{B}_y & -i \frac{1}{2} \kappa \mathbf{B}_x & 0 & 0
\end{array}\right).\label{eq:K-split}
\end{equation}
Physically speaking,  $l$-independent $K_{0}$ describes the graviton-photon mixing system free of external field, whereas $\delta K(l)$ is a perturbation matrix that takes
into account the interaction of system with the magnetic background. 
One can arrange Eq. (\ref{eom-U}) into $
\partial_{l}\left(e^{-i\int_{l_{0}}^{l}dl'K_{0}(l')}\mathcal{U}\left(l,l_{0}\right)\right)  =  ie^{-i\int_{l_{0}}^{l}dl'K_{0}(l')}\delta K(l)\mathcal{U}\left(l,l_{0}\right)
$
and iteratively solve it up to the first order
\begin{equation}
\mathcal{U}\left(l,l_{0}\right)  =  e^{i\left(l-l_{0}\right)K_{0}}+ie^{i\left(l-l_{0}\right)K_{0}}\int_{l_{0}}^{l}dl'e^{-i\left(l'-l_{0}\right)K_{0}}\delta K\left(l'\right)e^{i\left(l'-l_{0}\right)K_{0}}+O\left(\delta K^{2}\right).
\end{equation}
 Accordingly, the absolute value of terms  $\mathcal{U}_{31}, \mathcal{U}_{32}, \mathcal{U}_{41}, \mathcal{U}_{42}$ can be obtained as
 \begin{equation}
 	\begin{aligned}
 	\left|\mathcal{U}_{31}\left(l, l_0\right)\right|^2 &=\left|\mathcal{U}_{42}\left(l, l_0\right)\right|^2 =\frac{1}{4} \kappa^2 \int_{l_0}^{l} d l_1 \int_{l_0}^{l} d l_2 e^{-2 i\left(l_1-l_2\right) l_{o s c0}^{-1}} \mathbf{B}_x\left(l_1\right) \mathbf{B}_x\left(l_2\right),\\
 	\left|\mathcal{U}_{32}\left(l, l_0\right)\right|^2 &=	\left|\mathcal{U}_{41}\left(l, l_0\right)\right|^2=\frac{1}{4} \kappa^2 \int_{l_0}^{l} d l_1 \int_{l_0}^{l} d l_2 e^{-2 i\left(l_1-l_2\right) l_{o s c0}^{-1}} \mathbf{B}_y\left(l_1\right) \mathbf{B}_y\left(l_2\right),
 	\end{aligned} \label{4U}
 \end{equation}
where  $l_{\mathrm{osc0}}=\frac{2}{\left|n_{\mathrm{p} 1}\right| \omega}$. Note that $e^{-2 i\left(l_1-l_2\right) l_{\text {osc }}^{-1}}$ is the same for above all four $\mathcal{U}$ components due to the negligence of  the polarization difference.
 The parameters of interests are $B \lesssim 10^{-9} \textrm{Gauss}$ and $\omega \gtrsim 10^6 \textrm{Hz}$, thus an  order-of-magnitude estimations
show  that $\kappa B\lesssim10^{-31}\textrm{m}^{-1}$ and $10^{-22}\textrm{m}^{-1}\lesssim|n_{\textrm{pl}}|\omega\lesssim10^{-13}\textrm{m}^{-1}$. It means that magnetic term can be safely neglected so that  expression $l_{\textrm{osc}}=2/\sqrt{\kappa^{2}(\mathbf{B}_x^2+\mathbf{B}_y^2)+n_{\textrm{pl}}^{2}\omega^{2}}$ reduces to  $l_{\textrm{osc0}}=2/(|n_{\textrm{pl}}|\omega)$.
In this way, we consistently use the
 same symbol $l_{\textrm{osc}}$ throughout the paper.

Let us consider a stochastic magnetic field generated in the primordial
Universe. The primordial magnetic field is typically modelled as a
statistically isotropic Gaussian distributed random field with a vanishing
average expectation $\left\langle \mathbf{B}_{i}(\mathbf{x})\right\rangle =0$
but a non-vanishing correlation function \citep{Durrer:2013pga} as
\begin{equation}
\left\langle \mathbf{B}_{i}(\mathbf{x})\mathbf{B}_{j}\left(\mathbf{x}'\right)\right\rangle   =  \frac{1}{(2\pi)^{3}}\int d^{3}ke^{i\mathbf{k}\cdot\left(\mathbf{x}^{\prime}-\mathbf{x}\right)}\left[\left(\delta_{ij}-\hat{\mathbf{k}}_{i}\hat{\mathbf{k}}_{j}\right)P_{B}(k)-i \epsilon_{i j m} \hat{\mathbf{k}}_m P_{a B}(k)\right],\label{eq:B-ps}
\end{equation}
where $\hat{\mathbf{k}}=\mathbf{k}/k$ and $\epsilon_{i j m}$ is the totally antisymmetric symbol.  The $P_{B}(k)$ and $P_{aB}(k)$ are the symmetric and antisymmetric parts of the power spectrum. For a given field configuration,
the total magnetic energy density is $\rho_{B}=\left\langle \mathbf{B}^{2}(\mathbf{x})\right\rangle /2=\int dkk^{2}P_{B}(k)/\left(2\pi^{2}\right)$,
the energy density per unit and per logarithm $k$-interval are
$d\rho_{B}/dk=\rho_{B}(k)=k^{2}P_{B}(k)/\left(2\pi^{2}\right)$ and
$d\rho_{B}/d\textrm{ln}k=k^{3}P_{B}(k)/\left(2\pi^{2}\right)$ respectively.
The root mean square $B\equiv \sqrt{\left\langle \mathbf{B}^{2}(\mathbf{x})\right\rangle }$,
is used as a measure of the average strength of
the magnetic field. Moreover, the field strength smoothed over a certain
region of size $\lambda$ corresponds to $B_{\lambda}=\frac{8\pi}{\lambda^{3}}P_{B}\left(2\pi/\lambda\right)$.
The coherence length of stochastic magnetic field is given by $\lambda_{B}=2\pi\rho_{B}^{-1}\int\rho_{B}(k)k^{-1}dk$
and the corresponding strength density is $B_{\lambda_{B}}=\frac{8\pi}{\lambda_{B}^{3}}P_{B}\left(2\pi/\lambda_{B}\right)$.
In general, $\lambda_{B}$ represents the scale at which most of the
power energy is concentrated (except the very red spectrum like the
scale invariant one), thus $B_{\lambda_{B}}$ can be used to normalize
the power spectrum $P_{B}(k).$ Alternatively, the root mean square
$B$ can also be used to normalize $P_{B}(k).$ These two normalization
approaches lead to similar results with a sightly difference by an order
of one.

From Eq. (\ref{eq:B-ps}), one can extract the transverse part 
\begin{equation}
		\left\langle\mathbf{B}_x\left(l\right) \mathbf{B}_x\left(l^{\prime }\right)+\mathbf{B}_y\left(l\right) \mathbf{B}_y\left(l^{ \prime}\right)\right\rangle 
		 =\frac{1}{(2 \pi)^3} \int d^3 k e^{i k \cos \theta\left(l^{\prime}-l\right)}\left(1+\cos ^2 \theta\right) P_B(k), \label{BxBy}
\end{equation}
where $\hat{\mathbf{k}}_x \hat{\mathbf{k}}_x+\hat{\mathbf{k}}_y \hat{\mathbf{k}}_y+\hat{\mathbf{k}}_z \hat{\mathbf{k}}_z=1$ is used. Here $\theta$ denotes the angle between wave-vector $\mathbf{k}$ of the external magnetic field and the $l$-direction of GW propagation. 
Substituting Eqs. (\ref{4U}) and (\ref{BxBy}) into Eq. (\ref{P}), we average
the conversion probability over all possible magnetic field configurations
and obtain the main formula in this paper:
\begin{equation}
	\begin{aligned}
\left\langle\mathcal{P}(\triangle l) \right\rangle
=&\frac{1}{4} \kappa^2 \int_{l_0}^{l} d l_1 \int_{l_0}^{l} d l_2 e^{-2 i\left(l_1-l_2\right) l_{o s c}^{-1}}\left\langle\mathbf{B}_x\left(l_1\right) \mathbf{B}_x\left(l_2\right)+\mathbf{B}_y\left(l_1\right) \mathbf{B}_y\left(l_2\right)\right\rangle \\
 = & \frac{\kappa^{2}}{8\pi^{2}}\int\frac{1}{k}P_{B}(k)dk\left\{ 2k+\frac{1}{\triangle l}\left(\textrm{sin}\left(\left(2l_{\mathrm{osc}}^{-1}-k\right)\triangle l\right)-\textrm{sin}\left(\left(2l_{\mathrm{osc}}^{-1}+k\right)\triangle l\right)\right)\right. \\
&  +2\frac{4l_{\mathrm{osc}}^{-2}+k^{2}}{2l_{\mathrm{osc}}^{-1}-k}\textrm{sin}^{2}\left(\frac{1}{2}\left(2l_{\mathrm{osc}}^{-1}-k\right)\triangle l\right)-2\frac{4l_{\mathrm{osc}}^{-2}+k^{2}}{2l_{\mathrm{osc}}^{-1}+k}\textrm{sin}^{2}\left(\frac{1}{2}\left(2l_{\mathrm{osc}}^{-1}+k\right)\triangle l\right) \\
&  +4l_{\mathrm{osc}}^{-1}\left(\textrm{Ci}(\left|\left(2l_{\mathrm{osc}}^{-1}+k\right)\triangle l\right|)-\textrm{Ci}(\left|\left(2l_{\mathrm{osc}}^{-1}-k\right)\triangle l\right|)+\textrm{ln}\left|\frac{2l_{\mathrm{osc}}^{-1}-k}{2l_{\mathrm{osc}}^{-1}+k}\right|\right)\\\
&  \left.+\triangle l\left(4l_{\textrm{osc}}^{-2}+k^{2}\right)\left(\textrm{Si}\left(\left(2l_{\mathrm{osc}}^{-1}+k\right)\triangle l\right)-\textrm{Si}\left(\left(2l_{\mathrm{osc}}^{-1}-k\right)\triangle l\right)\right)\right\} ,\label{eq:P-stochastic}
 \end{aligned}
\end{equation}
where $\textrm{Ci}(x)=-\int_{x}^{\infty}\frac{\textrm{cos}(x')}{x'}dx'$
and $\textrm{Si}(x)=\int_{0}^{x}\frac{\textrm{sin}(x')}{x'}dx'$.
In the second equality we have used $\int_{l_{0}}^{l}dl^{\prime}\int_{l_{0}}^{l}dl''e^{i2\alpha\left(l''-l'\right)}=\frac{1}{\alpha^{2}}\sin^{2}(\alpha(l-l_{0}))$.
It should be pointed out that $k=2l_{\mathrm{\textrm{osc}}}^{-1}$ is
not a singularity. Moreover, the integral remains finite because of
the existence of physical infrared and ultraviolet cutoffs of $P_{B}(k)$.

\section{Resonance}

\label{sec:Resonance}

\begin{figure}
\includegraphics[scale=0.85]{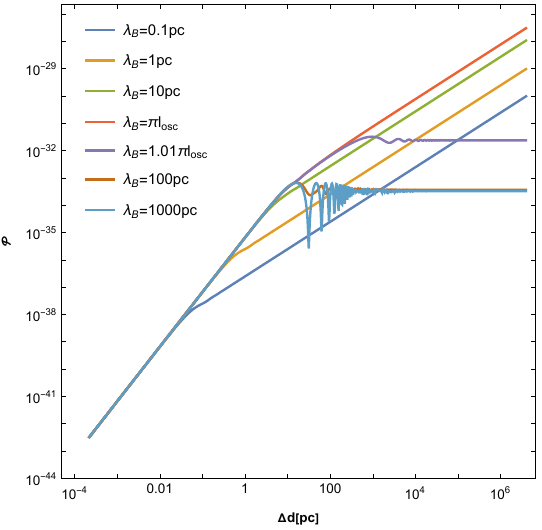}\includegraphics[scale=0.85]{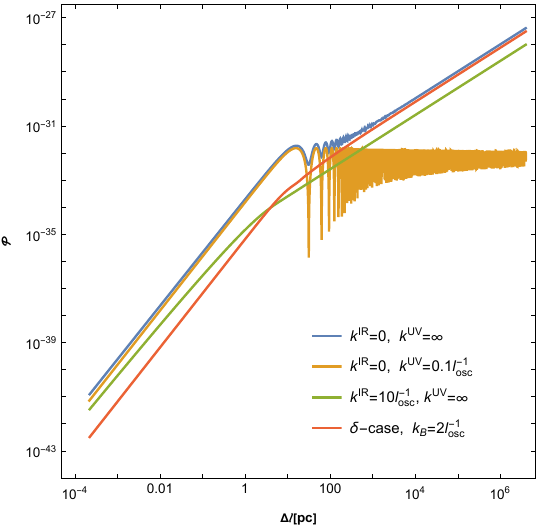}

\begin{tabular}{|c|c|c|c|}
\hline 
\multirow{4}{*}{monochromatic power spectrum} & \multirow{2}{*}{$\lambda_{B}>l_{osc}$} & $\triangle l\gtrsim l_{osc}$ & $\triangle l<l_{osc}$\tabularnewline
\cline{3-4} \cline{4-4} 
 &  & $\mathcal{P}\simeq\kappa^{2}B^{2}l_{osc}^{2}$ & $\mathcal{P}\simeq\kappa^{2}B^{2}\triangle l^{2}$\tabularnewline
\cline{2-4} \cline{3-4} \cline{4-4} 
 & \multirow{2}{*}{$\lambda_{B}\lesssim l_{osc}$} & $\triangle l\gtrsim\lambda_{B}$ & $\triangle l<\lambda_{B}$\tabularnewline
\cline{3-4} \cline{4-4} 
 &  & $\mathcal{P}\simeq\kappa^{2}B^{2}\lambda_{B}\triangle l$ & $\mathcal{P}\simeq\kappa^{2}B^{2}\triangle l^{2}$\tabularnewline
\hline 
\multirow{6}{*}{scale invariant power spectrum} & \multirow{2}{*}{$k^{IR}\lesssim l_{osc}^{-1}\lesssim k^{UV}$} & $\triangle l\gtrsim l_{osc}$ & $\triangle l<l_{osc}$\tabularnewline
\cline{3-4} \cline{4-4} 
 &  & $\mathcal{P}\simeq\kappa^{2}B^{2}l_{osc}\triangle l$ & $\mathcal{P}\simeq\kappa^{2}B^{2}\triangle l^{2}$\tabularnewline
\cline{2-4} \cline{3-4} \cline{4-4} 
 & \multirow{2}{*}{$l_{osc}^{-1}<k^{IR}<k^{UV}$} & $\triangle l\gtrsim1/k^{IR}$ & $\triangle l<1/k^{IR}$\tabularnewline
\cline{3-4} \cline{4-4} 
 &  & $\mathcal{P}\simeq\kappa^{2}B^{2}\triangle l/k^{IR}$ & $\mathcal{P}\simeq\kappa^{2}B^{2}\triangle l^{2}$\tabularnewline
\cline{2-4} \cline{3-4} \cline{4-4} 
 & \multirow{2}{*}{$k^{IR}<k^{UV}<l_{osc}^{-1}$} & $\triangle l\gtrsim l_{osc}$ & $\triangle l<l_{osc}$\tabularnewline
\cline{3-4} \cline{4-4} 
 &  & $\mathcal{P}\simeq\kappa^{2}B^{2}l_{osc}^{2}$ & $\mathcal{P}\simeq\kappa^{2}B^{2}\triangle l^{2}$\tabularnewline
\hline 
\end{tabular}

\caption{Upper: the conversion probability from graviton to photon in the presence
of the stochastic magnetic field with the monochromatic (Left) and
scale invariant (Right) power spectra. The characteristic oscillation
length of graviton-photon mixing is chosen to be $l_{\textrm{osc}}=10\textrm{pc}$
and the strength of the magnetic field is set to $B=10^{-12}\textrm{Gauss}$.
Lower: the pattern of conversion probability after GW traveling a
distance $\triangle l$ in two spectra. In upper plots, an order-of-magnitude estimation
is performed focusing on envelops
of the curves. Here conventions as $a>b$, $a\simeq b$
and $a<b$ correspond to $a>\mathcal{O}(10)b$, $a\simeq\mathcal{O}(1)b$
and $a<\mathcal{O}(0.1)b$, respectively.}

\label{fig-resonance}
\end{figure}

In last section we derived a general expression, Eq. (\ref{eq:P-stochastic}),
to calculate the probability of a graviton transition to a photon
in a stochastic magnetic background with spectrum $P_{B}(k)$. In order to
explore the dependence of conversion probability to traveling distance,
 oscillation length and magnetic spectrum, let us first consider
the magnetic spectrum as monochromatic-like. Specifically, we parametrize
the power spectrum as
\begin{eqnarray}
P_{B}(k) & = & \pi^{2}B^{2}\delta(k-k_{B})/k_{B}^{2}\label{eq:PB-delta}
\end{eqnarray}
with $k_{B}=2\pi/\lambda_{B}$. It corresponds to $d\rho/d\textrm{ln}k=B^{2}\delta(k/k_{B}-1)/2$,
which implies that all the energy is stored at a plane wave with a
certain wave-number $k_{B}$. In this specific case, Eq. (\ref{eq:P-stochastic})
can be  simplified. After performing a straightforward
calculation, we find that at sufficiently short distance, the probability
behaves as $\mathcal{P}\simeq\kappa^{2}B^{2}\triangle l^{2}$. At
large distance, the behaviour around $k_{B}=2l_{\mathrm{osc}}^{-1}$, corresponding to $\lambda_{B}=\pi l_{\textrm{osc}}$, plays a leading role. 
The last term in the bracket in Eq. (\ref{eq:P-stochastic}) is the most
relevant, corresponding to $\mathcal{P}\simeq\triangle l\left(4l_{\textrm{osc}}^{-2}+k_{B}^{2}\right)\left(\textrm{Si}\left(\left(2l_{\textrm{osc}}^{-1}+k_{B}\right)\triangle l\right)-\textrm{Si}\left(\left(2l_{\textrm{osc}}^{-1}-k_{B}\right)\triangle l\right)\right).$
For $k_{B}<2l_{\textrm{osc}}^{-1}$, i.e. $\lambda_{B}>\pi l_{\textrm{osc}}$,
this term does not diverge and thus the entire expression converges 
to a finite value $\mathcal{P}\simeq\kappa^{2}B^{2}l_{\textrm{osc}}^{2}$
for large enough $\triangle l$ and $\lambda_{B}$. Whereas $k_{B}\geq2l_{\textrm{osc}}^{-1}$,
i.e. $\lambda_{B}\leq\pi l_{\textrm{osc}}$, this term is proportional
to the $\triangle l$ and dominates over other terms, leading to an
enhanced probability $\mathcal{P}\simeq\kappa^{2}B^{2}\lambda_{B}\triangle l$
at large distance. Such an amplification effect is of particular interest. 
In order to visualize this effect more clearly, we set $l_{\textrm{osc}}=10\textrm{pc}$
and $B=10^{-12}\textrm{Gauss}$ and we plot the conversion probability as
a function of $\triangle l$ with different $\lambda_{B}$ in the
upper-left panel of Fig. \ref{fig-resonance}. In the case with $\lambda_{B}\gtrsim\mathcal{O}(10)l_{\textrm{osc}}$,
the probability curve first grows as $\mathcal{P}\propto\triangle l^{2}$
at short distance $\triangle l<l_{\textrm{osc}}$, then it oscillates 
at $\triangle l\simeq l_{\textrm{osc}}$ and finally
it reaches a plateau at $\mathcal{P}\simeq\kappa^{2}B^{2}l_{\textrm{osc}}^{2}$
when $\triangle l$ is sufficiently large. On the other hand, when $\lambda_{B}\leq\pi l_{\textrm{osc}}$,
the probability curve behaves similarly to the previous case at short distance $\triangle l\lesssim\lambda_{B}$,
but it increases linearly at $\triangle l\gtrsim\lambda_{B}$ besides reaching a constant value.
Such a linear amplification becomes strongest when a critical equality condition 
 $\lambda_{B}=\pi l_{\textrm{osc}}$ is satisfied  (the red line in upper-left
panel in Fig. \ref{fig-resonance}). It is worth mentioning that
a similar linear resonance effect has been observed in previous works \citep{carlson1994photon,Domcke:2020yzq,carenza2021turbulent}.
In particular, in Ref. \citep{carlson1994photon} the authors worked
in the context of electron density perturbation; in Ref. \citep{carenza2021turbulent}
this linear relation between probability and distance was found in
the $\lambda_{B}\ll l_{\textrm{osc}}$ case; we will discuss Ref.
\citep{Domcke:2020yzq} in more detail in the next section.

By using the monochromatic power spectrum $P_{B}\sim\delta(k-k_{i})$
as a basis, we can approximate any form of the multi-chromatic power
spectrum as $P_{B}\sim\sum_{i}\alpha_{i}\delta(k-k_{i})$ with varying
magnitudes $\alpha_{i}$. This approach allows us to infer the behaviour
of the conversion probability with a complicated power spectrum from
the well-studied monochromatic case. As an illustrative example, let
us consider the scale invariant power spectrum given by
\begin{eqnarray}
P_{B}(k) & = & \pi^{2}B^{2}/k^{3},\qquad k^{IR}<k<k^{UV},\label{eq:scaleinv-PBB}
\end{eqnarray}
where $k^{IR}$ and $k^{UV}$ are the infrared and ultraviolet
cutoffs, respectively. Because of scale invariance, the magnetic field strength
is smoothly averaged over a region of any size $\lambda$ within the
cutoffs, resulting in $B_{\lambda}=B$. In terms of energy density,
the monochromatic power spectrum Eq. (\ref{eq:PB-delta}) corresponds
to $d\rho/d\textrm{ln}k=B^{2}\delta(k/k_{B}-1)/2$, whereas the scale
invariant one is $d\rho/d\textrm{ln}k=B^{2}/2$ at $k^{IR}<k<k^{UV}$.
Thus, the scale-invariant power spectrum can be regarded as an opposite
extreme profile to the monochromatic scenario since its
energy is uniformly distributed across all scales. Due to the critical
point $k_{B}=2l_{\textrm{osc}}^{-1}$, its
relative position with respect to the cutoffs of scale invariant power
spectrum determines the behavior of probability. Indeed, as shown
in the upper-right panel of Fig. \ref{fig-resonance}, the pattern
of the probability curves are determined by the relative scales of
$l_{\textrm{osc}}^{-1}$, $k^{IR}$ and $k^{UV}$. When $k^{UV}<2l_{\textrm{osc}}^{-1}$,
the dependence of probability on distance is similar to the non-resonant
case in monochromatic scenario. When $2l_{\textrm{osc}}^{-1}\leq k^{UV}$,
the probability curve has a linear amplification in the large
distance. For $k^{IR}\leq2l_{\textrm{osc}}^{-1}\leq k^{UV}$,
the probability is maximally resonant (the critical red line in Fig.
\ref{fig-resonance}) and becomes insensitive to the cutoffs. This
can be easily understood if we roughly consider the scale invariant power
spectrum as a sum $P_{B}\sim\sum_{k_{i}}\delta(k-k_{i})$ with $k_{i}$
quasi-continues within the cutoff: then the main contribution comes
from the critical point, especially at sufficiently large distances.

\begin{figure}
\begin{centering}
\includegraphics{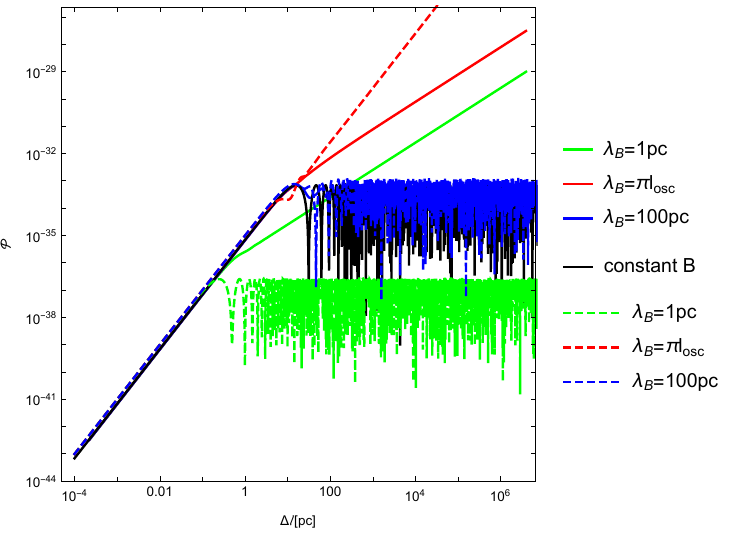}
\par\end{centering}
\caption{The conversion probability in different representative cases. The parameters are
chosen as $l_{\textrm{osc}}=10\textrm{pc}$ and $B=10^{-12}\textrm{Gauss}$.
The black line denotes the case with constant magnetic field (Eq.
(\ref{eq:P-constantB}) with $B= \sqrt {\mathbf{B}_x+\mathbf{B}_y}$). The green, red and blue solid lines represent
the monochromatic spectrum of the stochastic magnetic field with different
coherent lengths $\lambda_{B}$ (Eq. (\ref{eq:P-stochastic})). The
dashed lines represent the oscillatory magnetic field in certain wavelengths displayed in caption
(Eq. (\ref{eq:OscB}) with $B=\mathbf{B}_x$).}
\label{fig-Osc-Sto}
\end{figure}

In fact, the resonance effect observed above is essentially of the
same origin as the resonance addressed in Ref. \citep{raffelt1988mixing}.
In Ref. \citep{raffelt1988mixing} the authors studied a magnetic
field oscillating at a certain frequency. This situation is reminiscent
of the monochromatic spectrum, as the Fourier transformation of $\delta(k-k_{B})$
corresponds to a plane wave $e^{ik_{B}l}$. However, there are remarkable differences
between our case and that in Ref. \citep{raffelt1988mixing}. In Ref.
\citep{raffelt1988mixing} the resonance only occurs at exact condition
$\lambda_{B}=\pi l_{\mathrm{\textrm{osc}}}$ and the probability
scales as $\mathcal{P}\propto\triangle l^{2}$. In contrast, in our
case the resonance occurs in remarkably broader region for all $\lambda_{B}\leq\pi l_{\mathrm{\textrm{osc}}}$
and the probability scales as $\mathcal{P}\propto\triangle l$ (see
Fig. \ref{fig-Osc-Sto}). The reason for these differences lies in
the dimensional degrees of freedom of the system under study.  Indeed,
the magnetic model considered in Ref. \citep{raffelt1988mixing} is a 1-dimensional system, where the oscillatory magnetic
field is parameterized as $\mathbf{B}_x(l)=B\textrm{cos}(k_{B}(l-l_{0}))$ and $\mathbf{B}_y=\mathbf{B}_z=0$. Using
Eqs. (\ref{P}) and (\ref{4U}), we obtain the conversion probability
\begin{eqnarray}
\left\langle \mathcal{P}(\triangle l) \right\rangle  & = & \frac{1}{4}\kappa^{2}B^{2}\frac{1}{(-4l_{\mathrm{\textrm{osc}}}^{-2}+k_{B}^{2})^{2}}\left\{ 4l_{\mathrm{\textrm{osc}}}^{-2}\cos^{2}(k_{B}\triangle l)+4l_{\mathrm{\textrm{osc}}}^{-2}+k_{B}^{2}\sin^{2}(k_{B}\triangle l)\right.\nonumber \\
 &  & \left.-8l_{\mathrm{\textrm{osc}}}^{-2}\cos(k_{B}\triangle l)\cos(2l_{\mathrm{\textrm{osc}}}^{-1}\triangle l)-4l_{\mathrm{\textrm{osc}}}^{-1}k_{B}\sin(2l_{\mathrm{\textrm{osc}}}^{-1}\triangle l)\sin(k_{B}\triangle l)\right\} .\label{eq:OscB}
\end{eqnarray}
In the critical point $k_{B}=2l_{\mathrm{\textrm{osc}}}^{-1}$, i.e.
$\lambda_{B}=2\pi/k_{B}=\pi l_{\mathrm{\textrm{osc}}}$, the resonance
is excited and the probability increases as $\mathcal{P}(\triangle l)\propto\triangle l^{2}$.
In our model, the magnetic vector randomly points in different direction
in 3D spatial space. In mathematical terms, this introduces a misalignment
between the magnetic vector direction and the GW propagation direction,
represented by the angle $\theta$ in $e^{i\left(2l_{\mathrm{\textrm{osc}}}^{-1}-k\textrm{cos}\theta\right)\left(l''-l'\right)}$
in Eq. (\ref{eq:P-stochastic}). Consequently, the critical point becomes
$k\textrm{cos}\theta\simeq l_{\mathrm{\textrm{osc}}}^{-1}$, resulting
in a broader resonance region at $\lambda_{B}\lesssim l_{\mathrm{\textrm{osc}}}$.
In physical picture, the resonance with $\lambda'_{B}\lesssim l_{\mathrm{\textrm{osc}}}$
towards the GW propagation direction can be viewed as a projection
of the resonance with $\lambda_{B}\simeq l_{\mathrm{\textrm{osc}}}$
in another direction, where the angle between these two directions
is $\theta$ and $\lambda'_{B}=\lambda_{B}\textrm{cos}\theta$.

\section{discretization scheme}

\label{sec:Discritization-scheme}

In the previous section, we showed that the
probability of graviton conversion to photon can be resonantly amplified
with linear growth, in limit of large propagation distance.  However, an extremely long duration of the resonance
is unlikely to happen in realistic situations due to several decoherence
factors, e.g. the inhomogeneity of the electron density. Accordingly,
an appropriate discretization scheme to make a division along the
line of sight is demanded. In our case, the main source of decoherence 
arises from the expansion of the Universe. The graviton-photon mixing
system described by Eq. (\ref{eq:eom1}) is in Minkowski spacetime. In order to
incorporate the expansion of the Universe, we now work in the comoving
frame by replacing $\partial_{l}$ with $\frac{1}{a}\partial_{l}$
in Eq. (\ref{eq:eom1}). Moreover, we focus on the conversion process of  mixing system during the Post-Recombination era. Thus, we can still use the arguments as mentioned  in above Eq. (\ref{eom2})  to reasonably neglect the Cotton-Mouton effect as well as Faraday Rotation when relevant quantities are properly scaled. Therefore the modified equation 
reads as follows:
$\partial_l (h_{\times},h_{+},A_x,A_y)^T=i K(a,l)(h_{\times},h_{+},A_x,A_y)^T$ with
\begin{equation}
K (a,l)=a \left(\begin{array}{cccc}
\omega(a) & 0 & -i \frac{1}{2} \kappa \mathbf{B}_x(a) & i \frac{1}{2} \kappa \mathbf{B}_y(a) \\
0 & \omega(a) & i \frac{1}{2} \kappa \mathbf{B}_y(a) & i \frac{1}{2} \kappa \mathbf{B}_x(a) \\
i \frac{1}{2} \kappa \mathbf{B}_x(a) & -i \frac{1}{2} \kappa \mathbf{B}_y(a) & \omega(a)\left(n_{\mathrm{pl}}(a)+1\right) & 0 \\
-i \frac{1}{2} \kappa \mathbf{B}_y(a) & -i \frac{1}{2} \kappa \mathbf{B}_x(a) & 0 & \omega(a)\left(n_{\mathrm{pl}}(a)+1\right)
\end{array}\right),\label{eq:eom-expansion}
\end{equation}
where $a$ denotes the scaling factor and $l$ the comoving distance.
Here, quantities are scaled as $\omega(a)=\omega_{0}/a$, $\mathbf{B}_{x}(a,l)=\mathbf{B}_{x0}(l)/a^{2}$, $\mathbf{B}_{y}(a,l)=\mathbf{B}_{y0}(l)/a^{2}$
and $n_{\textrm{pl}}(a)=-e^{2}n_{b0}X_{e}(a)/(2a\omega_{0}^{2}m_{e})$,
where the subscript $0$ denotes the corresponding value at present
day, $n_{b0}\simeq0.25\textrm{m}^{-3}$ is the baryon number density
today and $X_{e}(a)$ is the ionization fraction. We consider 
 the magnetic field as diluted by the expansion of the Universe,
neglecting its dynamical evolution \citep{Durrer:2013pga}. Additionally,
we sharpen the parameterization as $X_{e}(a)\simeq10^{-4}$ ($0.002\lesssim a\lesssim0.05$)
and otherwise $X_{e}(a)\simeq1$ \citep{Kunze:2015noa}. As we will
see later, this simplification is enough accurate for our purposes.

Let us note that the non-commutativity of $K(a)$ at different cosmic time
renders searches of solutions for Eq.  (\ref{eq:eom-expansion}) more difficult than previous cases studied above. 
Therefore, we consider  a steady approximation as follows. We consider an interval
of cosmic time during which the relative change of $K$ due to the
Universe expansion has to be small enough, allowing us to
approximate the equation with a constant $K$ scaled at that specific
time. Physically, this approximation means that the GW does not ``feel''
the expansion of Universe when it propagates during such a short time
interval. We stress that one can not take the interval infinitely
small, otherwise the coherence of the system would be lost and hence
the total conversion probability along the line of sight would vanish.
In order to determine the appropriate time interval, we examine how $K$ changes
with respect to the conformal time $\tau$. We only need to concern
about the first three components of $K$ those scales as $1/a$. During
a time interval $\triangle\tau$, for example for the first component, it
relatively changes as $\triangle K_{11}/K_{11}=\mathcal{H}\triangle\tau,$
by using $d(1/a)/d\tau=\mathcal{H}/a$ with the comoving Hubble parameter
$\mathcal{H}$. The steady approximation demands that relative change
$\epsilon=\triangle K_{11}/K_{11}$ is small all the time. Regarding
the comoving distance $\triangle l$, this requirement can be translated
into $\triangle l=\triangle\tau=\epsilon\mathcal{H}^{-1}$. This means
that the GW traveling distance is a small portion of the comoving
Hubble radius $\mathcal{H}^{-1}$. This makes sense because the Hubble
radius characterizes the size of the local inertial frame in the expanding
Universe. In other words, the graviton-photon mixing system does not strongly "feel" the Universe expansion  in a suitable interval of  cosmic time.  A similar approximation has also been considered for axion-photon mixing in expanding Universe \citep{Higaki:2013qka}.

In terms of the redshfit $z=1/(1+a)$, the steady approximation leads
to a discretization scheme by setting a redshift sequence $[z_{1},z_{2},\cdots,z_{N}]$
iterated via $z_{n+1}=(1-\epsilon)z_{n}$ ($n=1,2,\cdots,N-1$). Since
we consider the primordial magnetic field at the Post-Recombination
era, the initial point is set to the end of Recombination $z_{1}\simeq1100$
and the iteration continues until the present day. The comoving distance
$\triangle l_{n}$ of GW path during interval patch $[z_{n},z_{n+1}]$
is given by $\triangle l_{n}=\epsilon\mathcal{H}^{-1}(z_{n})$, where
$\mathcal{H}(z_{n})=H_{0}(1+z_{n})^{1/2}$ with $H_{0}$ being the
Hubble parameter today. In each patch, we construct the kernel matrix
$K(z,x)$ with values evaluated at $z=z_{n}$ in Eq. (\ref{eq:eom-expansion})
and  perform the same perturbative approach as the Minkowski spacetime case. Thus, we obtain:{\footnotesize
\begin{equation}
\left\langle \mathcal{P}(\triangle l,z_{n}) \right\rangle   =  \frac{1}{\left(1+z_{n}\right)^{2}}\frac{1}{4}\kappa^{2}\int_{l_{0}}^{l}dl'\int_{l_{0}}^{l}dl''e^{-2i(l'-l'')l_{\textrm{osc}}^{-1}(z_{n})} \left\langle \mathbf{B}_{x}(z_{n},l')\mathbf{B}_{x}(z_{n},l'')+\mathbf{B}_{y}(z_{n},l')\mathbf{B}_{y}(z_{n},l'') \right\rangle, \label{P-expansion1}
\end{equation}}where $l_{\textrm{osc}}(z_{n})=2\left(1+z_{n}\right)/(|n_{\textrm{pl}}|\omega)$
is the comoving oscillation length. Here we use the same symbol
$l_{\textrm{osc}}$  as  clearly understood
 in both the static and expanding cases. To derive Eq. (\ref{P-expansion1}), we have used Eq. (\ref{P}) which is correct only when the initial GW mode is unpolarized. In fact, the GW can be perfectly regarded as an unpolarized source during the whole Post-Recombination
 	era  because of the suppressed conversion probability from GW to photon even in the occurrence of the resonance (typically $< 10^{-10}$, as we will see later). Finally, we obtain
the conversion probability during interval patch $[z_{n},z_{n+1}]$
under the expansion of universe as{\footnotesize{}
\begin{eqnarray}
\left\langle \mathcal{P}^{\textrm{exp}}(\triangle l_{n}) \right\rangle  & = & \frac{1}{\left(1+z_{n}\right)^{2}}\frac{\kappa^{2}}{8\pi^{2}}\int\frac{1}{k}P_{B}(z_{n},k)dk\left\{ 2k+\frac{1}{\triangle l_{n}}\left(\textrm{sin}\left(\left(2l_{\mathrm{osc}}^{-1}-k\right)\triangle l_{n}\right)-\textrm{sin}\left(\left(2l_{\mathrm{osc}}^{-1}+k\right)\triangle l_{n}\right)\right)\right.\nonumber \\
 &  & +2\frac{4l_{\mathrm{osc}}^{-2}+k^{2}}{2l_{\mathrm{osc}}^{-1}-k}\textrm{sin}^{2}\left(\frac{1}{2}\left(2l_{\mathrm{osc}}^{-1}-k\right)\triangle l_{n}\right)-2\frac{4l_{\mathrm{osc}}^{-2}+k^{2}}{2l_{\mathrm{osc}}^{-1}+k}\textrm{sin}^{2}\left(\frac{1}{2}\left(2l_{\mathrm{osc}}^{-1}+k\right)\triangle l_{n}\right)\nonumber \\
 &  & +4l_{\mathrm{osc}}^{-1}\left(\textrm{Ci}(\left|\left(2l_{\mathrm{osc}}^{-1}+k\right)\triangle l_{n}\right|)-\textrm{Ci}(\left|\left(2l_{\mathrm{osc}}^{-1}-k\right)\triangle l_{n}\right|)+\textrm{ln}\left|\frac{2l_{\mathrm{osc}}^{-1}-k}{2l_{\mathrm{osc}}^{-1}+k}\right|\right)\nonumber \\
 &  & \left.+\triangle l_{n}\left(4l_{\mathrm{osc}}^{-2}+k^{2}\right)\left(\textrm{Si}\left(\left(2l_{\mathrm{osc}}^{-1}+k\right)\triangle l_{n}\right)-\textrm{Si}\left(\left(2l_{\mathrm{osc}}^{-1}-k\right)\triangle l_{n}\right)\right)\right\} .\label{eq:P-expansion}
\end{eqnarray}
} Here the magnetic field spectrum $P_B(z_n,k)$ is defined in the same formula as in Eq. (\ref{eq:B-ps}) but with respect to the time-dependent correlation function $\left\langle\mathbf{B}_i(z_n,\mathbf{x}) \mathbf{B}_j\left(z_n,\mathbf{x}^{\prime}\right)\right\rangle$. The total conversion probability along the line of sight is given
by adding all probability contributions from all patches, namely $\mathcal{P}_{\textrm{total}}=\sum_{n=1}^{N-1}\mathcal{P}^{\textrm{exp}}(\triangle l_{n})$.

\begin{figure}
\includegraphics[scale=0.6]{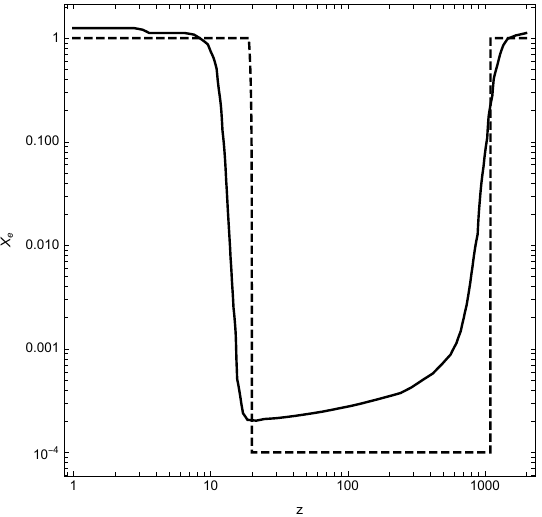}\includegraphics[scale=0.6]{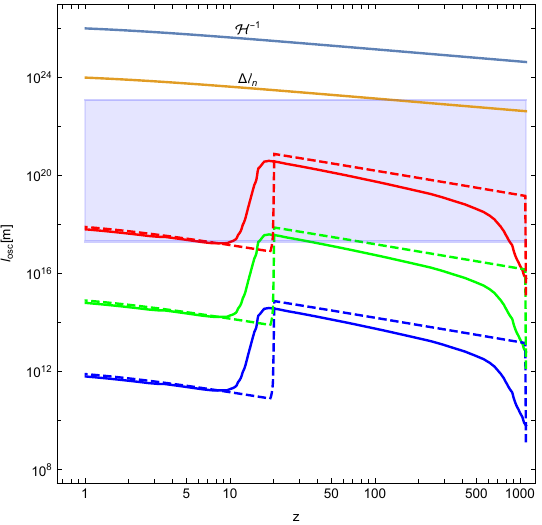}\includegraphics[scale=0.6]{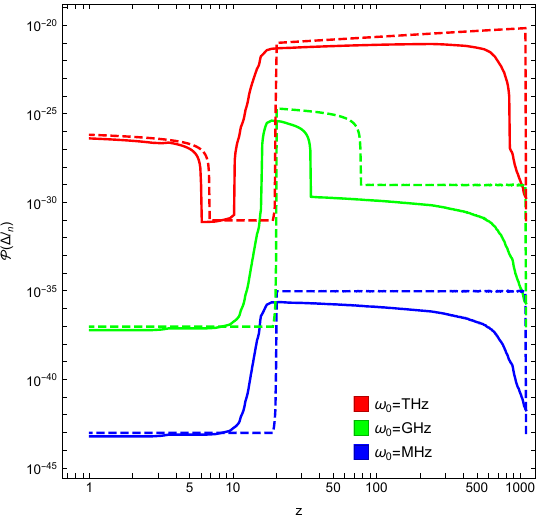}

\caption{Ionization fraction $X_{e}$, oscillation probability $l_{osc}$ and conversion probability versus the redshift in the Post-Recombination era.
The blue, green, red lines denote the different frequency cases with
$\omega_{0}=10^{6}$, $10^{9}$ and $10^{12}$Hz, respectively. The
solid line denotes an accurate parameterization of ionization fraction
function \citep{Kunze:2015noa}, whereas the dashed line denotes the
simplified parameterization in step shape. In the right panel, we
show the conversion probability $\mathcal{P}(\triangle l_{n})$ at
each redshift interval $[z_{n},z_{n+1}]$ with $\epsilon=0.01$. Here
we parameterize the magnetic field spectrum Eq. (\ref{eq:PB-real})
with  $n_{L}=-3$, $n_{S}=-4$, $k^{IR}=10^{-23}$, $k_{0}=10^{-18}$,
$k_{D}=10k_{0}$ and $A$ is normalized to the root mean square strength
$B=10^{-12}\textrm{Gauss}$. The shaded region in the middle panel
is bounded by $(2/k_{D},2/k^{IR})$.}
\label{fig1}
\end{figure}

In order to justify the aforementioned assumption of simplified parameterization
of ionization fraction $X_{e}(z)$, we show the oscillation length
$l_{\mathrm{\textrm{osc}}}(z)$ and conversion probability $\mathcal{P}^{\textrm{exp}}(\triangle l_{n})$
as a function of redshift in Fig. \ref{fig1}. As shown in this Figure, a step-like approximation of 
$X_{e}(z)$ does not relevantly change the main results. 
Moreover, this also shows that the conversion probability can
be safely neglected during the reionization era. The oscillation length
$l_{\mathrm{\textrm{osc}}}$ reaches its maximum value before the
reionization at $z\simeq20$, indicating that the corresponding
conversion probability is expected to be at its peak before reionization as well. However,
due to the expansion of the universe, the Eq. (\ref{eq:P-expansion})
includes an extra $(1+z)^{2}$ factor, which leads to a relative increase
in the probability at large redshifts. Nevertheless, during the Dark
Age $20\lesssim z\lesssim1000$, the relative changes of $l_{\mathrm{\textrm{osc}}}(z)$
as well as $\mathcal{P}^{\textrm{exp}}(\triangle l_{n})$ remains
within $\mathcal{O}(10)$. Therefore, using relevant quantities at
$z=20$ is sufficient to perform order-of-magnitude analysis. In addition,
the steep increase of the probability curve at $20\lesssim z\lesssim30$
for the case $\omega_{0}=\textrm{GHz}$ is caused by  the resonance when the condition $1/k_{D}\lesssim l_{\mathrm{\textrm{osc}}}\lesssim1/k^{IR}$
is satisfied (compare the green line and the shade region).

The last ingredient yet to be addressed in Eq. (\ref{eq:P-expansion})
is the power spectrum of stochastic magnetic field in the expanding
Universe. As mentioned in Introduction, cosmological phase transition
and inflation magnetogenesis are two main hypothetical mechanisms to
explain the generation of the magnetic field in the primordial universe
\citep{Durrer:2003ja,Durrer:2013pga}. In the case of cosmological
phase transition, the magnetic field is produced by bubbles coalescence
in a first order phase transition. This process leads to a blue spectrum
$P_{B}\sim k^{2}$ as a consequence of causality. On the other hand,
inflation magnetogenesis usually leads to a blue spectrum $P_{B}\sim k$.
To obtain a scale-invariant spectrum $P_{B}\sim k^{-3}$ during inflation,
 the inflaton has to be strongly coupled with vector fields and back-reactions are not negligible  \citep{Demozzi:2009fu}.
Nevertheless, several possible ways to solve these problems have
been explored in Ref. \citep{Subramanian:2015lua} and reference therein.

In the Post-Recombination era, we consider an evolving magnetic
field with the power spectrum that can be modelled as two power laws
\citep{Kahniashvili:2016bkp,Brandenburg:2018ptt}:
\begin{eqnarray}
P_{B}(z,k) & = & (1+z)^{4}P_{B_{0}}(k)=(1+z)^{4}\begin{cases}
A\left(k/k_{0}\right)^{n_{L}} & \text{ }k^{IR}<k\leq k_{0},\\
A\left(k/k_{0}\right)^{n_{S}} & \text{ }k_{0}<k<k_{D},
\end{cases}\label{eq:PB-real}
\end{eqnarray}
where $P_{B_{0}}(k)$ is defined with respect to $B_{0}(x)$ in the
comoving frame as Eq. (\ref{eq:B-ps}). Here, the $n_{L}$ is spectral
index at large length scale encoding informations about the magnetogenesis
mechanism. We focus on two representative cases
with $n_{L}=2$ and $n_{L}=-3$, corresponding to phase transition
and inflation magnetogenesis, respectively. For $n_{L}=2$ case, the
infrared cutoff $\text{ }k^{IR}$ is normally set to the Hubble radius
at the phase transition time. One can safely take $\text{ }k^{IR}$
to a infinite small value in practice since most of the power spectrum accumulates 
towards $k_{0}$. Whereas for $n_{L}=-3$ case, $\text{ }k^{IR}$
can be interpreted as the scale when magnetogenesis starts during
inflation \citep{Kahniashvili:2016bkp}. Its value could cover a large
range either being larger or smaller than the current Hubble radius.
On the other hand, at small scale the field is processed by magnetohydrodynamic
turbulence, leading to a universal Kolmogorov slope $n_{S}=-11/3$.
Here we introduce a weak turbulence with $n_{S}=-4$ as a different
MHD spectrum would only change the result by a numerical factor of
order unity. At smaller scale the spectral energy is damped away via
the viscosity of charged plasma during recombination, which is characterized
by the ultraviolet cutoff $k_{D}\simeq\mathcal{O}(100)(10^{-9}\textrm{Gauss}/B_{0})\textrm{Mp\ensuremath{c^{-1}}}$\citep{Kahniashvili:2009qi,Kahniashvili_2010}.
This cutoff defines a characteristic damping scale $\lambda_{D}\equiv2\pi/k_{D}$.
In addition, the factor $A$ is normalized by $B_{0}$.

\section{Numerical results}

\begin{figure}
\includegraphics[scale=0.78]{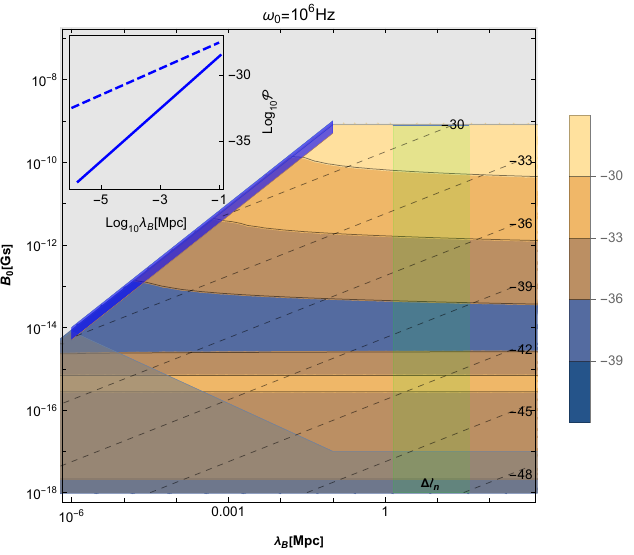}\includegraphics[scale=0.78]{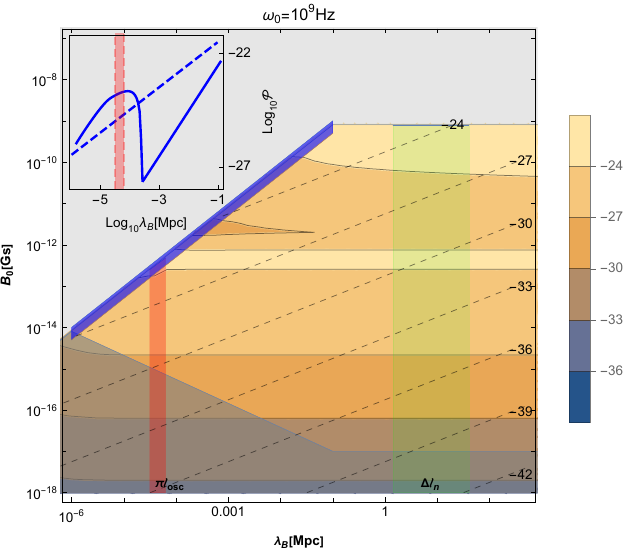}

\includegraphics[scale=0.78]{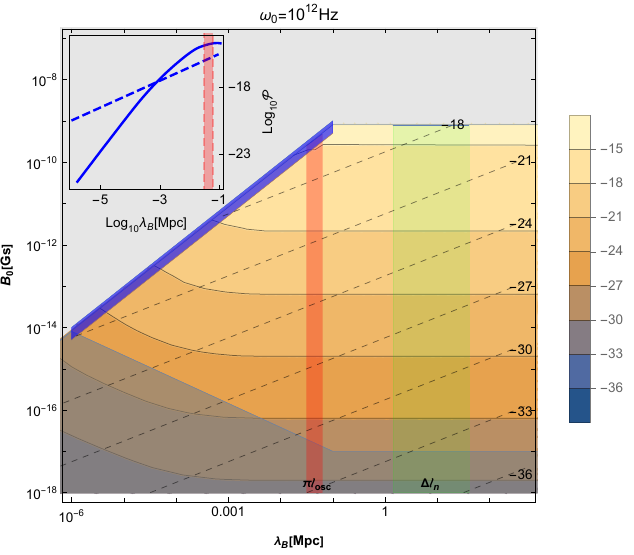}\includegraphics[scale=0.78]{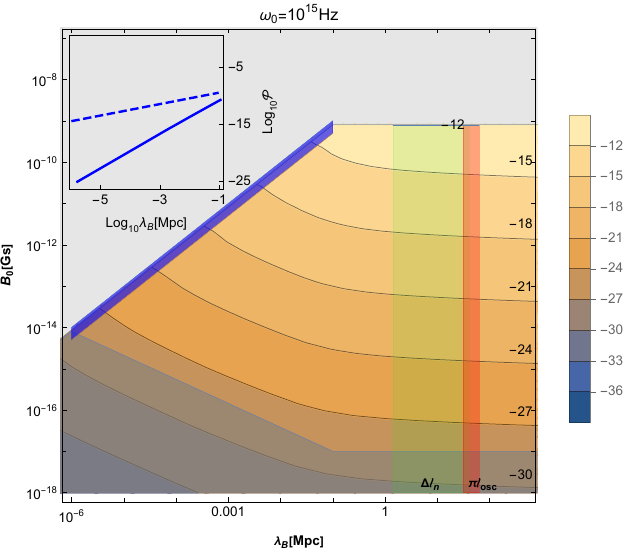}

\caption{The total probability $\log_{10}\mathcal{P}_{\textrm{total}}$ conversion
from graviton with ultrahigh frequency $\omega_{0}$ to photon propagating within the Post-Recombination era.
The gray region shows the exclusion on the primordial magnetic relic
as discussed in the text. The relic field from phase transition magnetogenesis
survives at the relatively smaller coherence scale today, as shown
in the narrow blue stripe region. The corresponding probability is
shown in color conventions displayed in captions. 
The inflationary magnetogenesis can generate a
scale invariant spectrum, with strength and coherence length varying
in a wide range. It could in principle fill the entire allowed observational
region given proper initial conditions. The red vertical line
denotes the typical oscillation length scale ($\sim10^{-8}\textrm{Mpc}$
in $\omega_{0}=10^{6}\textrm{Hz}$). The green area shows the rough
scale of the patch size in steady state discretization scheme. The
results in Ref. \citep{Domcke:2020yzq} are labeled in dashed lines.
Here we fix $\epsilon=0.01$.}

\label{fig-conversion-P}
\end{figure}

In this section we consider Eq. (\ref{eq:P-expansion}) and the aforementioned
discretization scheme. We will show numerical results of the total conversion
probability as the GWs travel through the Post-Recombination era in
the expanding Universe. The GWs in the ultrahigh frequency window
$\omega_{0}\geq\textrm{MHz}$ are of particular interests: they could
be potentially probed in radio channel through GW-photon conversion
processes. Therefore, we consider four typical frequency windows of
GW $\omega_{0}=10^{6}\textrm{Hz}$, $10^{9}\textrm{Hz}$, $10^{12}\textrm{Hz}$
and $10^{15}\textrm{Hz}$ and we study graviton-photon transitions in presence of the primordial magnetic background
field. The results are shown in Fig. \ref{fig-conversion-P}. The
gray region in $B_{0}$-$\lambda_{B}$ parameter plane is ruled out by CMB analysis and the magnetohydrodynamic turbulence
\citep{Durrer:2013pga}. Relic fields, lower than the limit from
 Blazar observations (the light gray zone), can not be directly
tested since they can be hidden by background from other dominant EM sources.
As elaborated in Ref. \citep{Durrer:2013pga}, the evolution of the
magnetic field generated during Electroweak or QCD phase transition
undergoes to the MHD process and it stops at the narrow blue region of parameters corresponding to Universe
today. Note that we ignore the reionization era which negligibly contributes
to the total probability. In contrast, the hypothetical scale invariant
magnetic field generated during inflation is more interesting from
the observational perspective because it could in principle be within 
the entire allowed observable region. Before discussing them separately,
we briefly introduce two characteristic scales. One is the scale of
the patch size $\triangle l_{n}\simeq\mathcal{O}(1-10)\textrm{Mpc}$
along the line of sight (the green shaded region in Fig. \ref{fig-conversion-P}). The other is the
oscillation length $l_{\textrm{osc}}(z_{n})$ of graviton-photon mixing,
which varies in the expanding Universe. Since the total conversion
probability receives the main contribution when the oscillation length
is at its maximum scale (see Fig. \ref{fig1}), we refer to it as
$l_{\textrm{osc}}$ in the following and label the corresponding value
in the red vertical line.

For the relic magnetic field arising from phase transition, its strength
and coherent length are constrained to the blue stripe region in Fig.
\ref{fig-conversion-P}. The spectrum modeled in Eq. (\ref{eq:PB-real})
has a blue scaling at small $k<k_{0}$ and red scaling at large
$k>k_{0}$. Thus, the profile of the power spectrum resembles a monochromatic
one, peaked around $k_{0}$, with the coherent length at scale $\lambda_{B}\sim k_{0}^{-1}$.
As a result, the resonant effect occurs when $\lambda_{B}\lesssim l_{\textrm{osc}}$
and it reaches its maximum at $\lambda_{B}\simeq l_{\textrm{osc}}$ (see
the left panel in Fig. \ref{fig-resonance}). Note
that we perform an order-of-magnitude analysis since the spectrum
is not a perfect $\delta$-shape but has finite width around $k_{0}$.
Nevertheless, the main result of resonance can be applied yet. Indeed,
in $\omega_{0}=10^{9,12,15}\textrm{Hz}$ cases, the conversion probability
is resonantly amplified when the coherent length is comparable to
the oscillation length scale (see the insets in Fig. \ref{fig-conversion-P}).

In contrast, the scale invariant relic magnetic field from inflation
is much less constrained and could fill the entire observable region.
In this scenario, $\lambda_{B}$ loses its original physical
meaning because the strength of magnetic field is equally distributed 
over all scales. In fact, the definition $\lambda_{B}\sim\int kP_{B}(k)dk$
indicates that for the scale invariant spectrum $P_{B}\sim k^{-3}$,
the value of $\lambda_{B}$ is mainly determined by the infrared cutoff,
namely $\lambda_{B}^{-1}\simeq k^{IR}$. Nonetheless, we still use
$\lambda_{B}$ for convenience for comparisons with observations. Generally,
as indicated in Fig. \ref{fig-conversion-P}, the conversion probability
increases as the GWs frequency $\omega_{0}$ or magnetic strength
$B_{0}$ increases. The total probability can arrive up to $10^{-12}$
when $\omega_{0}=10^{15}\textrm{Hz}$ and $B_{0}\sim10^{-9}\textrm{Gauss}$.
Such a sizable conversion probability could have detectable cosmological
implications which deserve for future dedicated studies. For a fixed strength
of magnetic field, the probability is insensitive to the coherent
length in relatively low frequency cases with $\omega_{0}=10^{6}\textrm{Hz}$
and $\omega_{0}=10^{9}\textrm{Hz}$. In relatively high frequency
cases with $\omega_{0}=10^{12}\textrm{Hz}$ and $\omega_{0}=10^{15}\textrm{Hz}$,
the probability increases with a larger coherence length of the magnetic
field. Regarding the resonance effect, the analysis in Sec. \ref{sec:Resonance}
reveals that the resonance is excited when the damping scale is comparable
or smaller than the oscillation length, i.e. $\lambda_{D}\lesssim l_{\textrm{osc}}$
(see the upper-right panel in Fig. \ref{fig-resonance}). In Fig.
\ref{fig-conversion-P}, one can observe that the conversion probability
generally decreases as the strength of the magnetic field decreases.
However, in $\omega_{0}=10^{6}\textrm{Hz}$ and $\omega_{0}=10^{9}\textrm{Hz}$
cases, the probability significantly increases by more than about
$3\sim6$ orders of magnitude when the resonance condition $\lambda_{D}\lesssim l_{\textrm{osc}}$
is satisfied. Moreover, the probability is maximally resonant in a
large region where the condition $\lambda_{D}\lesssim l_{\textrm{osc}}\lesssim\lambda_{B}$
is satisfied. This results in an amplification of the probability
for the relatively weaker magnetic field and the field with relatively
larger coherence length.

In Tab. \ref{table-compare}, we list the patterns of the probability
in different parameter regions by the order-of-magnitude analysis.
Here we omit all details and we only focus on distinct features for with/without
resonance. We can see that the probability for the scale invariant
scenario has much richer structure than the phase transition
one. Such distinct features of the probabilities can be used
to probe these two magnetogenesis scenarios from early Universe.

\begin{table}
{\small{}}%
\begin{tabular}{|c|c|c|c|}
\hline 
\multirow{4}{*}{{\footnotesize{}phase transition}} & \multirow{1}{*}{{\footnotesize{}$\omega_{0}=10^{6}\textrm{Hz}$($\triangle l_{n}>l_{\textrm{osc}}$,$\triangle l_{n}>\lambda_{B}$)}} & {\footnotesize{}$\lambda_{B}>l_{\textrm{osc}}$} & {\footnotesize{}$\mathcal{P}_{\textrm{total}}\simeq\kappa^{2}B^{2}l_{\textrm{osc}}^{2}D/\triangle l_{n}$}\tabularnewline
\cline{2-4} \cline{3-4} \cline{4-4} 
 & \multirow{2}{*}{{\footnotesize{}$\omega_{0}=10^{9}\textrm{Hz}$($\triangle l_{n}>l_{\textrm{osc}}$,$\triangle l_{n}>\lambda_{B}$)}} & {\footnotesize{}$\lambda_{B}>l_{\textrm{osc}}$} & {\footnotesize{}$\mathcal{P}_{\textrm{total}}\simeq\kappa^{2}B^{2}l_{\textrm{osc}}^{2}D/\triangle l_{n}$}\tabularnewline
\cline{3-4} \cline{4-4} 
 &  & {\footnotesize{}$\lambda_{B}\lesssim l_{\textrm{osc}}$} & {\footnotesize{}$\mathcal{P}_{\textrm{total}}\simeq\kappa^{2}B^{2}\lambda_{B}D$}\tabularnewline
\cline{2-4} \cline{3-4} \cline{4-4} 
 & {\footnotesize{}$\omega_{0}=10^{12,15}\textrm{Hz}$($\triangle l_{n}>l_{\textrm{osc}}$,$\triangle l_{n}>\lambda_{B}$)} & {\footnotesize{}$\lambda_{B}\lesssim l_{\textrm{osc}}$} & {\footnotesize{}$\mathcal{P}_{\textrm{total}}\simeq\kappa^{2}B^{2}\lambda_{B}D$}\tabularnewline
\hline 
\multirow{11}{*}{{\footnotesize{}\makecell{scale invariant \\ (inflation)}}} & \multirow{2}{*}{{\footnotesize{}$\omega_{0}=10^{6}\textrm{Hz}$($\triangle l_{n}>l_{\textrm{osc}}$,$\lambda_{B}>l_{\textrm{osc}}$)}} & {\footnotesize{}$\lambda_{D}\lesssim l_{\textrm{osc}}\lesssim\lambda_{B}$} & {\footnotesize{}$\mathcal{\mathcal{P}_{\textrm{total}}}\simeq\kappa^{2}B^{2}l_{\textrm{osc}}D$}\tabularnewline
\cline{3-4} \cline{4-4} 
 &  & {\footnotesize{}$l_{\textrm{osc}}<\lambda_{D}<\lambda_{B}$} & {\footnotesize{}$\mathcal{\mathcal{P}_{\textrm{total}}}\simeq\kappa^{2}B^{2}l_{\textrm{osc}}^{2}D/\triangle l_{n}$}\tabularnewline
\cline{2-4} \cline{3-4} \cline{4-4} 
 & \multirow{4}{*}{{\footnotesize{}$\omega_{0}=10^{9}\textrm{Hz}$($\triangle l_{n}>l_{\textrm{osc}}$)}} & {\footnotesize{}$\lambda_{D}\lesssim l_{\textrm{osc}}\lesssim\lambda_{B}$} & {\footnotesize{}$\mathcal{\mathcal{P}_{\textrm{total}}}\simeq\kappa^{2}B^{2}l_{\textrm{osc}}D$}\tabularnewline
\cline{3-4} \cline{4-4} 
 &  & \multirow{2}{*}{{\footnotesize{}$\lambda_{D}<\lambda_{B}<l_{\textrm{osc}}$}} & {\footnotesize{}$\mathcal{\mathcal{P}_{\textrm{total}}}\simeq\kappa^{2}B^{2}\lambda_{B}D$($\triangle l_{n}\gtrsim\lambda_{B}$)}\tabularnewline
\cline{4-4} 
 &  &  & {\footnotesize{}$\mathcal{\mathcal{P}_{\textrm{total}}}\simeq\kappa^{2}B^{2}\triangle l_{n}D$($\triangle l_{n}<\lambda_{B}$)}\tabularnewline
\cline{3-4} \cline{4-4} 
 &  & {\footnotesize{}$l_{\textrm{osc}}<\lambda_{D}<\lambda_{B}$} & {\footnotesize{}$\mathcal{\mathcal{P}_{\textrm{total}}}\simeq\kappa^{2}B^{2}l_{\textrm{osc}}^{2}D/\triangle l_{n}$}\tabularnewline
\cline{2-4} \cline{3-4} \cline{4-4} 
 & \multirow{2}{*}{{\footnotesize{}$\omega_{0}=10^{12}\textrm{Hz}$($\triangle l_{n}>l_{\textrm{osc}}$,$\lambda_{D}<l_{\textrm{osc}}$)}} & {\footnotesize{}$\lambda_{D}\lesssim l_{\textrm{osc}}\lesssim\lambda_{B}$} & {\footnotesize{}$\mathcal{\mathcal{P}_{\textrm{total}}}\simeq\kappa^{2}B^{2}l_{\textrm{osc}}D$}\tabularnewline
\cline{3-4} \cline{4-4} 
 &  & {\footnotesize{}$\lambda_{D}<\lambda_{B}<l_{\textrm{osc}}$} & {\footnotesize{}$\mathcal{\mathcal{P}_{\textrm{total}}}\simeq\kappa^{2}B^{2}\lambda_{B}D$}\tabularnewline
\cline{2-4} \cline{3-4} \cline{4-4} 
 & \multirow{3}{*}{{\footnotesize{}$\omega_{0}=10^{15}\textrm{Hz}$($\triangle l_{n}<l_{\textrm{osc}}$,$\lambda_{D}<l_{\textrm{osc}}$)}} & {\footnotesize{}$\lambda_{D}\lesssim l_{\textrm{osc}}\lesssim\lambda_{B}$} & {\footnotesize{}$\mathcal{\mathcal{P}_{\textrm{total}}}\simeq\kappa^{2}B^{2}\triangle l_{n}D$}\tabularnewline
\cline{3-4} \cline{4-4} 
 &  & \multirow{2}{*}{{\footnotesize{}$\lambda_{D}<\lambda_{B}<l_{\textrm{osc}}$}} & {\footnotesize{}$\mathcal{\mathcal{P}_{\textrm{total}}}\simeq\kappa^{2}B^{2}\lambda_{B}D$($\triangle l_{n}\gtrsim\lambda_{B}$)}\tabularnewline
\cline{4-4} 
 &  &  & {\footnotesize{}$\mathcal{\mathcal{P}_{\textrm{total}}}\simeq\kappa^{2}B^{2}\triangle l_{n}D$($\triangle l_{n}<\lambda_{B}$)}\tabularnewline
\hline 
\multirow{2}{*}{{\footnotesize{}domain-like model}} & \multirow{2}{*}{{\footnotesize{}all $\omega_{0}$($\lambda_{B}=\triangle l_{n}$)}} & {\footnotesize{}$\lambda_{B}>l_{\textrm{osc}}$} & {\footnotesize{}$\mathcal{\mathcal{P}_{\textrm{total}}}\simeq\kappa^{2}B^{2}l_{\textrm{osc}}^{2}D/\lambda_{B}$}\tabularnewline
\cline{3-4} \cline{4-4} 
 &  & {\footnotesize{}$\lambda_{B}\lesssim l_{\textrm{osc}}$} & {\footnotesize{}$\mathcal{\mathcal{P}_{\textrm{total}}}\simeq\kappa^{2}B^{2}\lambda_{B}D$}\tabularnewline
\hline 
\end{tabular}{\small\par}

\caption{All possible cases of the total conversion probability after traveling a
total distance $D$ in different parameter regions are displayed. We perform order-of-magnitude
estimations (see Fig. \ref{fig-resonance}). The parameter conditions (see Fig.
\ref{fig-conversion-P}) are shown in the parentheses in the second
column while corresponding probabilities in the third column.}
\label{table-compare}
\end{table}

Let us  comment on the existing domain-like magnetic field model as well
as its variations, which have been adopted in almost all relevant
literature dealing with cosmic magnetic fields. In these models, the
line-of-sight region is divided into many patches, each with a size
equal to the coherence length $\lambda_{B}$. To realize the stochastic
property of magnetic field, in each patch the magnetic field is assumed
to be uniform but chosen in a random direction. This scheme provides several simplifications
for probability computations 
 and it is often sufficient to capture the main features of the
system. However, this approach is somehow artificial and it violates
divergence free condition of magnetic field. On the contrary,
our approach is based in statistical techniques and it 
is more realistic and natural to describe the stochastic configuration
of magnetic field. Moreover, our approach is capable of integrating
any kinds of discretization scheme. It means that the size of discrete
patches is not necessary identified with coherence length
of magnetic field. For a useful comparison, we add the pattern of the
probability derived in the domain-like model in Tab. \ref{table-compare}.
We can see that the results obtained  from our approach is remarkably different to ones from domain-like model due to the resonance effect
and the discretization scheme. It is interesting to note that, in a particular case with $\lambda_{B}\simeq\triangle l_{n}$,
the phase transition scenario in our method leads to the same result
of the domain-like model. Whereas in the scale invariant scenario,
when $\lambda_{D}\lesssim l_{\textrm{osc}}$ and $\lambda_{B}\simeq\triangle l_{n}>l_{\textrm{osc}}$
our result receives a maximal resonant contribution, which is larger
than the domain-like model by a factor $\lambda_{B}/l_{\textrm{osc}}$.
This factor can be as large as about $10^{9}$ at $\omega_{0}=10^{6}\textrm{Hz}$,
$10^{5}$ at $\omega_{0}=10^{9}\textrm{Hz}$ and $10^{2}$ at $\omega_{0}=10^{12}\textrm{Hz}$.

We stress that even if the authors of Ref. \citep{Domcke:2020yzq}
also perform a perturbative approach, their results are completely
different from our ones. The conversion probabilities obtained in Ref.
\citep{Domcke:2020yzq} (the Fig. 2 therein) are shown in dashed lines
in Fig. \ref{fig-conversion-P}. In Ref. \citep{Domcke:2020yzq},
the conversion rate is given by $\varGamma\sim\kappa^{2}B^{2}l_{\textrm{osc}}^{2}/\lambda_{B}$,
and hence the total probability is approximated as $\mathcal{\mathcal{P}_{\textrm{total}}}\simeq\varGamma D\simeq\kappa^{2}B^{2}l_{\textrm{osc}}^{2}D/\lambda_{B}$.
Therefore, the model in Ref. \citep{Domcke:2020yzq} effectively resembles
a domain-like model under the condition $\lambda_{B}=\triangle l_{n}\gtrsim l_{\textrm{osc}}$
(see Tab. \ref{table-compare}). Furthermore, in Ref. \citep{Domcke:2020yzq}
the authors applied their probability formula to impose upper bounds
on the stochastic GW background derived from EDGES \citep{Bowman:2018yin}
and ARCADE 2 \citep{fixsen2011arcade} experiments at $\omega_{0}\sim10^{8}\textrm{Hz}$
and $10^{10}\textrm{Hz}$, respectively. For the scale
invariant case with $\omega_{0}=10^{9}\textrm{Hz}$ in Fig. \ref{fig-conversion-P},
we can see that the maximum probability in the entire viable parameter
region is comparable to the one in Ref. \citep{Domcke:2020yzq}.
However, 
the remarkably different distribution of the probability density indicates
that in several subregions of $B_{0}$-$\lambda_{B}$ plane, the corresponding
bounds on GWs obtained in Ref. \citep{Domcke:2020yzq} can be either
overestimated or underestimated by several orders of magnitude.
In other words, comparative analysis with experimental data are model-dependent and previous bounds have to be revisited in case of resonances,
especially in case of scale-invariant power spectrum. 
For
instance, at $B_{0}\lesssim10^{-12}\textrm{Gauss}$ or $\lambda_{B}\gtrsim\textrm{Mpc}$,
the conversion probability predicted from our model is $5\sim10$ orders larger,
hence lowering the corresponding GW bounds by $2\sim5$ orders. From
the reversed perspective,  radio signals could probe the
magnetic field with much weaker strength or larger coherence length
than previously considered

\begin{figure}
	
\includegraphics[scale=0.5]{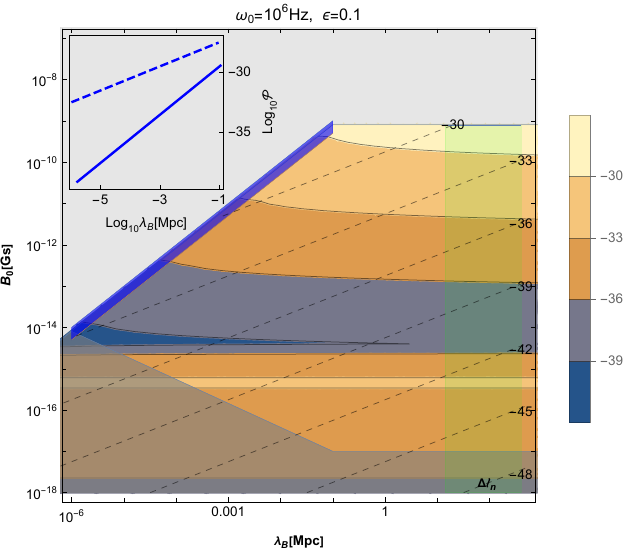}
\includegraphics[scale=0.5]{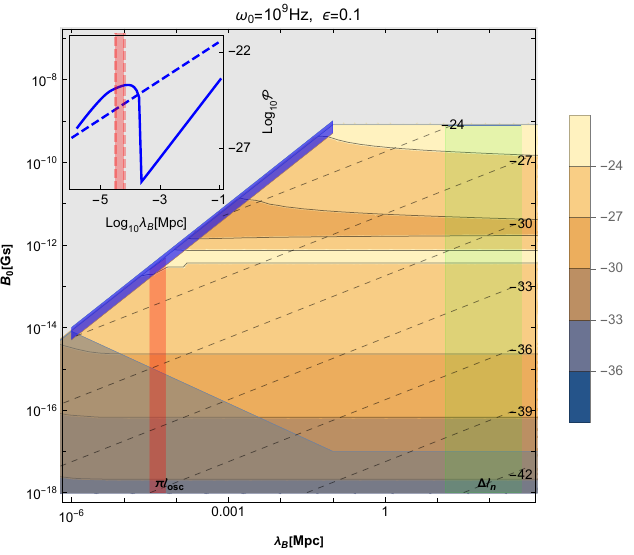}\includegraphics[scale=0.5]{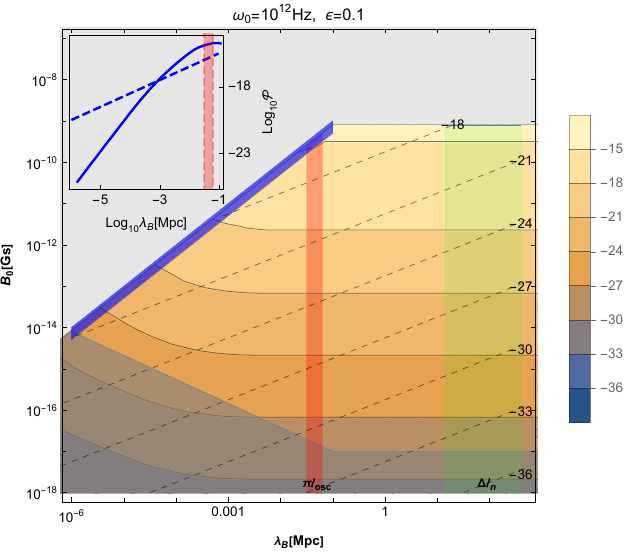}

\includegraphics[scale=0.5]{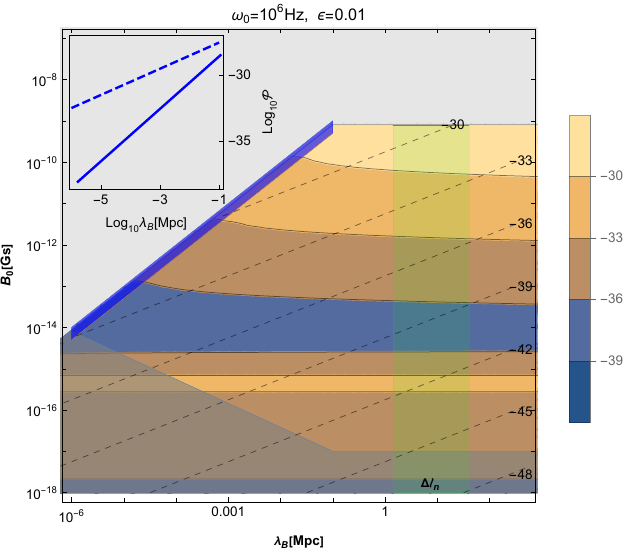}\includegraphics[scale=0.5]{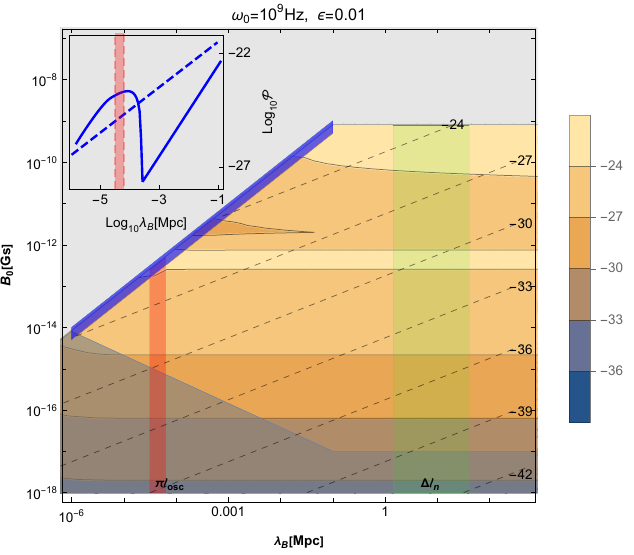}\includegraphics[scale=0.5]{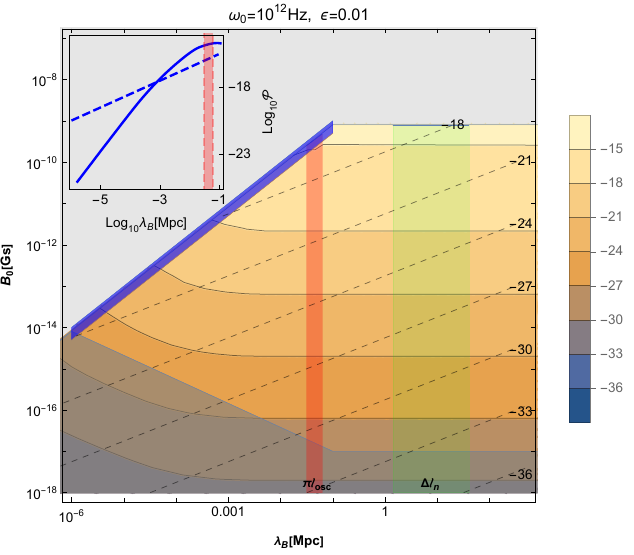}

\includegraphics[scale=0.5]{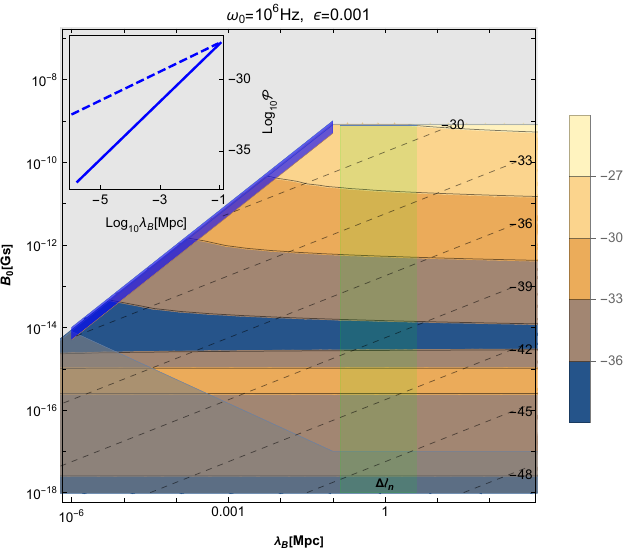}\includegraphics[scale=0.5]{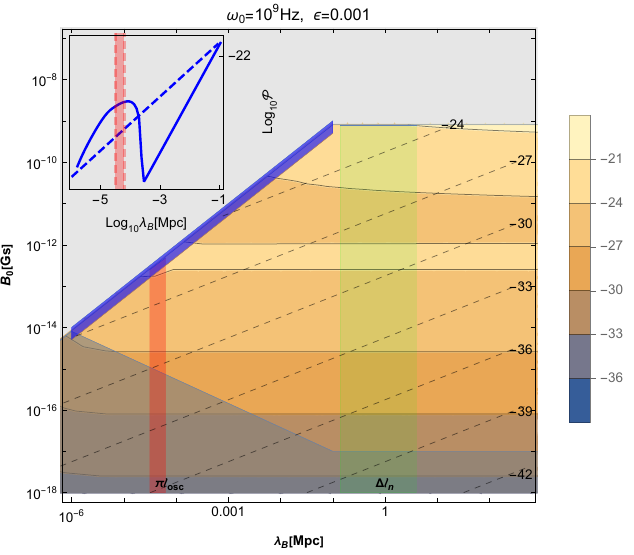}\includegraphics[scale=0.5]{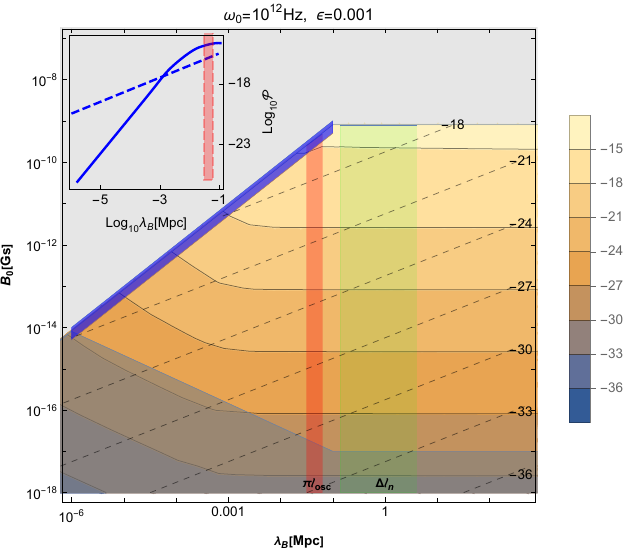}

\caption{The total conversion probability $\textrm{log}_{10}\mathcal{P}_{\textrm{total}}$
in different $\epsilon=0.1$, $0.01$ and $0.001$ cases. The panels
in the second row with $\epsilon=0.01$ are same to Fig. \ref{fig-conversion-P}.
The dashed lines denote the conversion probability obtained in Ref.
\citep{Domcke:2020yzq}.}

\label{fig-different-sigma}
\end{figure}

In the end, we mention that a different choice for the value of $\epsilon$, different 
than $0.01$ assumed above, would affect the results as setting
a different discretization. Nevertheless, one can always perform
similar analysis observing the pattern of the probability behaviour
in different parameter regions as illustrated in Fig. \ref{fig-resonance}.
For any choice of $\epsilon$, the behaviour of probability in our model would significantly
differ from the one in domain-like model.  In order to demonstrate the $\epsilon$-dependence, we show  three cases with $\epsilon=0.1$, $0.01$
and $0.001$ in Fig. \ref{fig-different-sigma}. The order-of-magnitude
estimation in the Tab. \ref{table-compare} can be applied to these
cases as well. It is worth noticing that when the resonance happens (see the THz case for instance), the total conversion probability is insensitive to the choice of $\epsilon$. This can be easily understood because, in the linear resonance region $\mathcal{P}(\triangle l_n) \varpropto \triangle l_n$, the total conversion probability, after a long fixed distance $D$, is expected to be $\mathcal{P}_{total}(D) \simeq N \mathcal{P}(\triangle l_n) \varpropto D  $ with $N\simeq D/\triangle l_n$ the number of  patches.  However, in the non-resonance region, a sensitive $\epsilon$-dependence of result 
reflects the fact that the
discretization scheme is crucial. 

\section{Discussion and Conclusions}

\label{sec:Discussion-and-Conclusion}

In summary, our study focused on the inverse Gertsenshtein effect
in a graviton-photon mixing system with an inhomogeneous stochastic
magnetic background. The probability of conversion from graviton to
photon can be obtained from a perturbative analysis. We found that the
conversion probability can be resonantly amplified in certain parametric spaces.
We first worked in the Minkowski spacetime and considered two simplified
representative spectra parameterized in monochromatic and scale invariant
forms. For the monochromatic one, the resonance is obtained when the coherence
length of the magnetic field $\lambda_{B}$ becomes comparable or
smaller than the oscillation length of the graviton-photon mixing
$l_{\textrm{osc}}$. The resonance effect is maximally amplified at
$\lambda_{B}\simeq l_{\textrm{osc}}$. On the other hand, for the scale invariant case, the resonance
band lies in the region where the damping scale $\lambda_{D}$ is
comparable or smaller than $l_{\textrm{osc}}$. 

Then, we included the expansion
of the Universe in our analysis:
cosmological acceleration 
 enters as a decoherence factor in the transition probability.  
In particular, we performed a steady
approximation by dividing the GW propagation distance into patches with each
size corresponding to  $\triangle l_{n}\simeq\mathcal{O}(1)\textrm{Mpc}$.
Concerning the primordial magnetic field, we focus on the phase transition and
inflation magnetogenesis. For the relic field generated from phase transition,
only a narrow region in the observational $B_{0}$-$\lambda_{B}$
plane is allowed. We showed that the corresponding conversion probability has peak at
$\lambda_{B}\simeq l_{\textrm{osc}}$ when the resonance has its max. 
For the scale invariant field generated during inflation,
its strength and the coherence length are allowed in a large range which can 
 fill the whole observational parameter space. In
such a magnetic background, the conversion from graviton to photon
is resonantly enhanced once the damping scale is smaller than the
oscillation length $\lambda_{D}\lesssim l_{\textrm{osc}}$. This
amplification can raise the conversion probability up to $2\sim5$ orders of magnitude with respect to domain-like models. Thus, the distinct features of the probability function profiles can potentially provide 
a way to distinguish inflation magnetogenesis from phase transitions one. 

Moreover, we made a comparison between our model with the domain-like
one and current experimental bounds. 
We found that the distribution
of the probability in $B_{0}$-$\lambda_{B}$ plane are different
to ones obtained in Ref. \citep{Domcke:2020yzq}. 
Thus, to include resonances leads to a revisit of GW bounds 
derived from
EDGES and ARCADE2 experiments in Ref. \citep{Domcke:2020yzq}
of several orders of magnitude, especially 
for small fields  $B_{0}\lesssim10^{-12}\textrm{Gauss}$
or large coherent lengths $\lambda_{B}\gtrsim\textrm{Mpc}$. Conversely, possible
radio signals can probe the cosmological magnetic field with much weaker magnetic strength
or much larger coherence length.

Let us end with some short remarks. The resonance phenomenon
is not exclusive to graviton-photon mixing but could manifest in other
mixing systems described by Eq. (\ref{eom2}) with an inhomengeneous
background field, such as axion-photon mixing.
Concerning the inter-galactic or intra-galactic magnetic
fields in the late Universe, in principle our approach can also be applied. However,
the estimation of the conversion probability strongly depends on the
specific model of galactic winds, magnetization process, volume filling
factor, ionization levels and other relevant effects.

\noindent \textbf{Acknowledgements.}

A.A.  is supported by 
National Science Foundation of China (NSFC) No.12350410358;
the Talent Scientific Research Program of
College of Physics, Sichuan University, Grant No.1082204112427 \&
the Fostering Program in Disciplines Possessing Novel Features for
Natural Science of Sichuan University, Grant No.2020SCUNL209 \& 1000
Talent program of Sichuan province 2021. S.C. acknowledges the support
of Istituto Nazionale di Fisica Nucleare (INFN) (iniziative specifiche MoonLight2 and 
QGSKY). This paper is based upon work from the COST Action CA21136,
\textit{Addressing observational tensions in cosmology with systematics
and fundamental physics} (CosmoVerse) supported by COST (European
Cooperation in Science and Technology)

\end{document}